\documentclass[10pt]{article}
\usepackage{graphicx}
\usepackage{float}
\usepackage{amsthm}
\usepackage{authblk}

\newcommand{\ket}[1]{|{#1} \rangle}

\renewcommand{\phi}{\varphi}
\newcommand{\dash}{\ensuremath{{}_{\mbox{--}}}}

\RequirePackage{amsmath}
\RequirePackage{amssymb} 

\usepackage{tabularx}

\date{\today}

\usepackage{hyperref}
\begin{document}

\title{Benchmarking quantum computers via protocols} 

\author[1]{Dekel Meirom}
\author[2]{Tal Mor}
\author[2]{Yossi Weinstein}

\affil[1]{Faculty of Electrical and Computer Engineering, Technion, Haifa, Israel}

\affil[2]{Computer Science Department, Technion, Haifa, Israel}

\maketitle

\section*{Abstract}
Benchmarking quantum computers often deals with the parameters
of single qubits or gates and sometimes deals with algorithms run
on an entire chip or a noisy simulator of a chip.

Here we propose the idea of using \emph{protocols} to benchmark
quantum computers. The advantage of using protocols, especially the 
seven suggested here, over other benchmarking methods 
is that there is a clear cutoff (i.e., a threshold)
distinguishing quantumness from classicality for each of our protocols.

The protocols we suggest enable a comparison among 
various circuit-based quantum computers,
and also between real chips and their noisy simulators. This latter
method may then be used to better understand the 
various types of noise of the real chips. 
We use some of these protocols to answer
an important question: ``How many \emph{effective qubits} are there
in this $N$-qubit quantum computer/simulator?'', and we then conclude
which \emph{effective sub-chips} can be named ``truly-quantum''.


\section{Introduction}\label{sec:intro}

Benchmarking quantum computers is a complicated task. Various methods are used,
such as a single parameter named \emph{quantum volume}\cite{quantum-volume}, a
method to characterize average gate error by using gates from the Clifford group named \emph{randomized
benchmarking}\cite{randomized-benchmarking}, a method to run complicated algorithms named \emph{algorithmic qubits}\cite{algorithmic-qubits} 
and other holistic methods \cite{circuit-mirroring}.
%
%
%
Different companies prefer to use different methods, potentially in order
to show the advantage or the success of their architecture and chips.

One natural benchmarking method that is universal but not simple, 
is to compare between different quantum computers in their ability   
to run quantum algorithms that perform various important tasks. 
In particular, it would be useful to be able 
to compare quantum computers in the
performance of tasks for which quantum algorithms are known or believed 
to perform better or even exponentially
better than classical ones, such as factoring large numbers. 
For example, the quantum Fourier transform has been used 
to compare and benchmark quantum computers~\cite{Monroe}. 
However, 
proving a \emph{practical} advantage of any specific quantum 
algorithm over the
classical one is challenging during the current NISQ era, 
where NISQ stands for
``noisy intermediate-scale quantum''.  
Can any currently existing quantum computer factorize in one day
all the numbers composite of two primes, $N=PQ$, where $P$ and $Q$ are 
prime numbers of just 5 or 6 bits? Due to the various types of noise,  
probably not yet. Clearly even if it is possible
or soon will become possible, it won't tell us much about the advantage
of quantum computers 
over classical ones, and will potentially tell us only a little 
about the advantage of
one quantum computer over another.

\subsection{Preliminaries: Protocols as a method for benchmarking}
\label{Protocols-are-good}
To make benchmarking simpler and more universal, we suggest using  
several \emph{quantum protocols}. 
Various quantum protocols have a clear and easily proven
advantage over classical ones. 
Therefore, quantumness can be easily
shown in theory and in practice for quantum protocols. 

By the term ``protocol"
we mean an algorithm (here --- rather simple algorithms) 
run by two or more parties that are at different
locations. The simplest example of a protocol is: The sender (Alice) 
sends a bit to the receiver (Bob),
and its quantum version: Alice sends a qubit to Bob. 
We notice that the distance
between Alice and Bob might have a major influence on the quality of the sent
information, when the information is quantum and is sensitive to interaction
with the environment. 

Here we adapt the term ``locations" to be
a qubit or a set of qubits in Alice's hands, and another qubit
or a set of qubits that are far apart in Bob's hands, \emph{on a quantum device}, which we also call \emph{a real chip}. 
Similarly --- these locations
may be far apart on a
simulator simulating the quantum device, which we also call
\emph{a fake-chip}, following IBM's names for simulators. 
When the statement is relevant for both the real chip and the fake chip,
we simply say ``the chip''.
Here we also adapt the term ``to send" not
to mean that there are photons running from Alice to Bob, 
but that we \emph{swap} one or more qubits from
Alice's location on the device to Bob's location on the device. Finally, 
we adapt the term ``distance" to mean the number of swap operations done
to reach from Alice's site to Bob's site.
We do not allow an overlap between Alice's site and Bob's site. 

When, along some minimal path
connecting Alice and Bob, there are $L'$ 
qubits in between Alice's site and Bob's
site, the distance is defined as $L=L'+1$. 
Therefore, if Alice's site is exactly 
adjacent to Bob's site ($L' = 0$) 
we consider the distance of the
protocol to be $L=1$. This means that exactly one swap operation 
is required for the transport of a qubit from Alice to Bob.
Internal swaps by Alice or internal swaps by Bob are not counted
when calculating the distance of the protocol.
The qubits between Alice's site and Bob's site
are named ancilla qubits, or ancillas.
In the current work we also name Bob's qubits \emph{ancillas}, because
in the seven specific protocols we present here they
are not modified --- Bob never initializes
any of the qubits originally in his site, nor uses them, 
and he does not care about their states.
Bob only uses qubits arriving from Alice.

For simplicity, and for generality, i.e. to easily fit any circuit-based 
quantum computer, our focus is on two-party protocols, hence there are 
just two
locations (or two nodes) --- Alice's node and Bob's node. 
If Alice's node or Bob's contains more than a single qubit, we always
define the node to contain connected qubits, namely a connected
substructure. 
More general protocols with more than just two parties can be defined, but this
is beyond the scope of the current work.




Quantum protocols are characterized by unique features. Two of these 
features are quite simple, and thus, from our point of view, 
lend themselves perfectly to 
a comparison between different quantum computers/simulators:

\begin{description} 
\item Given a \textbf{single qubit} in an unknown state, 
sending it or teleporting it
from Alice to Bob can be done theoretically with
fidelity one, and in practice with some fidelity between zero and one,
where a fidelity of half means the qubit became totally randomized. 
However, if Alice communicates only classically, Bob can 
regenerate the state of that qubit with fidelity at
most $2/3$, using the results Alice sent him after measuring the qubit~\cite{quantumness-threshold}. 
Thus we consider a fidelity of at most two thirds as a proof for
losing the quantumness of the protocol, and a fidelity above $2/3$ as a success
--- the protocol is still \emph{quantum}.

\item 
Given a \emph{pair of maximally entangled qubits} one qubit 
in Alice's hands and one
in Bob's hands, their entanglement can theoretically reach fidelity 1, but in
practice suffers from some reduction of the fidelity, where a fidelity
of $1/4$ for each Bell state 
means the two qubits could simply be at random states. 
Furthermore, 
disturbed entanglement of fidelity at most 1/2 can be generated via local
operations and classical communication between Alice  
and Bob at the two 
locations~\cite{werner-states}. More precisely, a density matrix 
which is a mixture of one Bell state with probability 1/2  
with the other three Bell states each with probability of $1/6$ can be generated
with no quantum communication at all but by local operations and classical
communication. And even 
a mixture of one Bell state with probability 1/2  
with the other three Bell states each with probability of $p_1$, $p_2$
and $p_3$ (namely $p_1+p_2+p_3 = 1/2$)  can be generated with no quantum
communication.

Here we adopt this threshold for the case where the final Bell state is in one
location:  if initially it is perfect at Alice's hands
(or supposedly perfect) yet
after some communication process it is imperfect at Alice's hands or at Bob's
hands, the threshold of 1/2 is our threshold as the measure of quantumness.   
I.e., we consider a fidelity of at most 1/2 as a proof for entirely  
losing the quantumness of the protocol, 
and a fidelity above $1/2$ as a success
--- the protocol is still \emph{quantum}.

\end{description}



We present here seven protocols. We  
use the first five protocols for defining 
the 
\emph{protocols vector}, a list of five quantum fidelities, which we find to be
important for comparing the abilities of
many current day quantum computers.  
This protocol vector is therefore a 5-number quality control measure.
Our protocol vector is useful for an entire real chip or fake chip
(a simulator). It is also useful  
in case quantumness is not obtained for a whole chip, 
but is obtained on a specific subchip.  


However, even if a quantum computer (either a real chip 
or a fake-chip, a whole chip or a whole subchip) 
is found to be quantum for all five protocols, 
it is not the end of the game. We designed
two generalized protocols that could most probably be much
more challenging for any chip and for various subchips. 

\subsection{Chips and effective subchips}
Whenever a device 
fails to be quantum for any specific protocol (or generalized
protocol) of the seven protocols presented here, 
we may define and find an effective ``subchip'' such
that for that subchip the protocol succeeds, between any two locations 
for Alice and Bob.
This method defines a smaller subchip and hence also a smaller number of 
\emph{effective qubits}. 
We often search for the exact borderline of showing quantumness, 
i.e., the protocol is still quantum for a distance $L$ but not for a distance
$L+1$.
Then we define, in some cases, a large (potentially 
the largest) effective subchip. 
This method may answer an important
question: ``How many effective qubits does this $N$-qubit computer have?''.

Each protocol can be used to compare two  
chips, or to find
the borderline of quantumness for a single chip, or to find 
an effective subchip for a single chip.
Furthermore, each protocol can indicate differences between
the real chip and its simulator, for a better understanding of various
noises.


In the most general case, for any specific 
``given'' locations of Alice's qubits
 and Bob's qubits, one might want to check all possible ways
of connecting Alice and Bob, and choose the one giving the highest fidelity.
However this would be far too cumbersome.
For all currently available architectures, 
a good approximation can be obtained by limiting the search for
the optimal path from Alice to Bob. We consider three options of limited
paths.

\begin{enumerate} 
\item 
The linear subset: In the extreme (yet highly useful) 
case we define a single line, such that Alice's
qubits, Bob's qubits, and the path between them are all along that single
line.  We check all possible lines in such a case, when we
search for the effective subchip. This is a very special case as it
limits Alice's substructure and Bob's substructure to be located on
a line.

Given a linear subset, let's order Alice's qubits 
and Bob qubits from left to right such that the leftmost qubit is Alice's
and the rightmost qubit is Bob's. 
If the connectivity of the chip is such that this is a shortest path between 
Alice's leftmost qubit and Bob's rightmost qubit,  we can call it ``a shortest linear subset''. 
In this paper, when we consider a linear subset, it is always the shortest.
\item
We mostly restrict
the communication between Alice and Bob 
to be along one line, namely ``the path'',
both in the case where Alice and Bob are located along a single 
line and in the case where
Alice and Bob have arbitrarily located substructures 
(as in Subsection~\ref{ssec:Cat-Kolkata}).
If a single qubit is sent, we identify the shortest paths (one or more)
and restrict our tests to these paths.
When two or more qubits are sent, they are 
sent via the same path (in the current case).


\item 
We do not demonstrate cases where two or more qubits are sent along different paths, with the exception of ion-trapped qubits.
In ion-trapped qubits the use of a single path is far from a good approximation for the optimal fidelity.

\end{enumerate} 

There is no single way to define optimal paths in a manner
that applies to any connectivity map. 
Therefore, we mostly consider two clear cases for optimal paths:
\begin{itemize}
\item
Whenever the chip or a subchip has low connectivity, 
it is possible to analyze
all ``shortest linear subsets'' for any two qubits. 
\item
Whenever the chip or a subchip is fully connected, it is possible to allow
using all L=1 paths between Alice and Bob.
\end{itemize}

\subsection{Organization of the paper}

The rest of this paper is organized as follows.
In Section~\ref{sec:protocols} 
we present the seven protocols for comparing
quantum computers:  Five basic protocols --- do-nothing, superdense coding,
Bell-state transfer, teleportation, and entanglement swapping, and two 
generalized protocols --- generalized do-nothing and cat state.
The five basic protocols and the two generalized
protocols are firstly fully presented 
on a single line, for simplicity:   
We actually present a simple practical case for each protocol, 
the case of six qubits on a line which includes Alices's and Bob's qubits. We perform this 
on a simulator named Fake-Kolkata, containing 27 qubits, 
that simulates a real chip of IBM-Q named Kolkata. 
The two generalized protocols are defined with some freedom, 
via the line connecting two line-shaped substructures,
one substructure at Alice's location and another, no larger, substructure at Bob's location. 

In Section~\ref{sec:results} we present our main results as follows:
First, we provide histograms for protocols~\ref{item:do-nothing},~\ref{item:superdense} and~\ref{item:ent-swap} and a few numerical results 
for all the six-qubit protocols defined and demonstrated in 
Section~\ref{sec:protocols}. 

Then, we focus on IBM's (15-qubit)  
Melbourne real chip and its simulator  
in Subsection~\ref{ssec:res-Melbourne}. We mainly
present our results on IBM's
simulator of Melbourne, due to much less availability
of the real Melbourne, 
focusing on the number of
effective qubits and presenting effective subchips, for specific 
protocols, and for the entire protocols vector --- we concluded 
that there are 9 effective qubits in the
effective simulated subchip. 
On the real chip we only ran various cases of the do-nothing protocol, achieving quantumness in some paths up to a length of 6,
while on the fake one all five basic protocols were checked.
We showed fidelity versus distance graphs, and we 
also compared the real Melbourne to its simulator.



In Subsection~\ref{ssec:res-Kolkata} we present our results on IBM's
simulator fake-Kolkata (27-qubit), obtaining positive
results for several of 
the five simpler protocols. For each of the simpler protocols,
the entire fake-chip is found to be effective.

In Subsection~\ref{ssec:Cat-Kolkata}, we run a generalized protocol on a structure
that goes beyond the linear subset.
For the cat-state protocol, 
we found the boundary between a success and a failure of presenting
quantumness.

In Subsection~\ref{ssec:checking-ancillas} 
we present the ``do-nothing'' protocol when ancilla qubits are non-zero 
to start with, and are inluded in the fidelity results. 

In Subsection~\ref{ssec:res-Eagle} we present a few results on 
IBM's Eagle chip and fake-chip (a family of 127-qubit chips):
We check a chip named Brisbane,
comparing the real chip and the fake-chip, 
and comparing both with fake-Kolkata. 


In Subsection~\ref{ssec:res-fake-IonQ} 
we present a few results on a
simplified simulators of Ion-Q's 11-qubit Harmony 
and 25-qubit Aria fake-chips.
We use a different path (each of $L=1$ distance) for each sent
qubit.

Finally, in Section~\ref{sec:Conclusion} we summarize our conclusion 
and present various important questions that can be answered in future work.

\section{Our protocols and a fully simulated subchip} \label{sec:protocols}

We present five basic protocols in which Alice's and Bob's number of qubits is
fixed per each protocol. These five protocols provide our main 
quality control measure --- the vector protocol, and our main conclusion
for a real chip or a simulator --- defining an effective subchip.
In addition, we present two generalized protocols, in which Alice's 
number of qubits (hence also Bob's) is flexible. 
These protocols can be especially
useful in the case where the entire chip is effective according to
the simpler five protocols, and yet, one wants
to test its limits. 

Three of our five basic protocols and one of the two generalized protocols
employ one-way quantum communication, and the remaining ones employ
two-way quantum communication.  In all the two-way protocols the ancillas
are back to their original locations by the end of the protocol, while
in all the one way protocols the ancillas are shifted towards Alice's location
by swaps, depending on the number of qubits sent to Bob. We always 
either let all the sent qubits stay at Bob's node, or return all of them 
 back to Alice, in our seven protocols.
The state of the ancilla qubits, whether shifted or not,
should not be modified by the protocol, 
so it may or may not be initially the state $\ket{0}$. 
In most of our demonstrations here the state of the ancilla qubits is zero at
the start of the protocol, and is supposed to be zero at the end as well. Yet,
we also clarify here and demonstrate 
in Section~\ref{ssec:checking-ancillas}, 
the case in which the state of the ancilla qubits, prior to running the
protocol, may be arbitrary. 


For each of the seven protocols below, the number in braces is the minimal number of qubits required for the protocol, when L = 1.
\begin{enumerate}
\item \label{item:do-nothing}A single qubit protocol named the do-nothing protocol, 
$\{2\}$. 
\item \label{item:superdense}A two-qubit protocol named the super-dense coding 
protocol~\cite{superdense},  
$\{3\}$. 
\item A two-qubit protocol named a Bell-state transfer protocol, 
$\{4\}$. 
\item A three-qubit protocol named the teleportation
protocol~\cite{teleportation}, 
$\{4\}$. 
\item \label{item:ent-swap}A four-qubit protocol named the entanglement swapping
protocol~\cite{ent-swap}, 
$\{6\}$. 

\item A $M$-qubit protocol named the generalized do-nothing protocol,
$\{2M\}$, where the simplest case of $M=1$ is in item number 1 above. 

\item A $\{M;J\}$-qubit protocol named the cat-state protocol,
$\{M+J\}$, where the simplest case of $M=J=2$ is in item number 3 above. 
\end{enumerate}

For simplicity, we ran all seven protocols on 6 qubits,
all on a single line,
on a simulator
named fake-Kolkata, that simulates the real IBM-Q
chip named Kolkata.  The Kolkata chip has 27
qubits.  
This is very a simple case of analyzing just
a single path between Alice and Bob, while Alice's and Bob's
substructures are trivial (namely along that same line).

We recall that in the protocols we use here, Bob never prepares any qubits. He always receives $n_B$ qubits, and either keeps them or sends 
them back to Alice, depending on the protocol. 
Hence, any qubits originally at Bob's site are considered ancillas.
In all the protocols, there is the option to initialize 
the ancilla qubits to some
arbitrary state via $U_{ancillas}$ at step 1. We mention
this option explicitly, but in parentheses,  
only for the first protocol. 
In that case, during the last steps of the protocol we will 
rotate the ancilla qubits back to the zero state using 
$U_{ancillas}^\dagger$, before
measuring them. 
For all other protocols, we do not apply operations to the ancilla qubits beside swap. Hence, we usually do not measure the ancilla qubits, excluding some examples later on. 

\begin{figure}[h]
    \centering
    \includegraphics[width=0.75\linewidth]{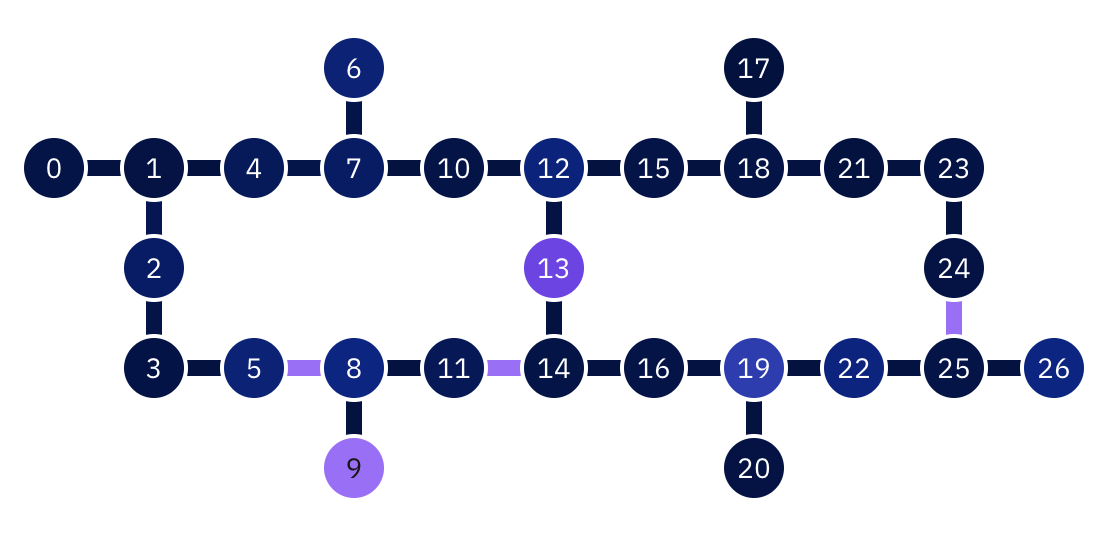}
    \caption{IBM-Q's Kolkata; colors of qubits or connections 
provide information about quality (that may change from week to week).}
    \label{fig:Kolkata}
\end{figure}

The Kolkata chip is shown in Figure~\ref{fig:Kolkata}.
%
In this paper, when we consider various paths,
we commonly describe a relevant path simply as the relevant sequence of 
qubits along the path, e.g., 
$q_0 \rightarrow q_1 \rightarrow q_2 \rightarrow q_3 \rightarrow q_5 \rightarrow q_8$.

\subsection{The seven protocols}

Here we explicitly present the seven protocols we suggest, where $n$
is the number of qubits involved in any run of a protocol. 
In parentheses we write the numbers $[ n_A ; n_B ]$ 
of the qubits in Alice's site (node) and in Bob's site (node),
where $n_A+n_B \le n \le N$. Where $N$ is the number of qubits in the chip. Note that 
$n=n_A+n_B$ 
is the minimal number of qubits for which the protocol is
applicable, 
and is relevant when Alice's node
is adjacent to Bob's node, with no ancillas in between.
Note also that 
$N = n = n_A+n_B$ is the minimal size of a chip for which
a specific protocol is still useful, 
e.g. from the following protocols, $2$ for the 
do-nothing, $3$ for superdense-coding, $6$ for entanglement swapping, 
and $M+J$ for the cat state protocols, 

\begin{enumerate}
\item The do-nothing protocol --- 
 $[1;1]$:
\begin{enumerate}
\item Reset the single work qubit, $q_0$, and the five ancilla qubits to zero.
[Optional: apply any unitary transformation $U_{ancillas}$ to all ancilla qubits.]
\item Alice initializes $q_0$ to a random state via a transformation
$U_{work}$.  
\item Apply swap gates to move $q_0$ to
Bob's node at the $I$'th location. 

In case we choose $I=q_8$, the distance (i.e., number of swaps) is $L=5$, and
the number of relevant ancilla qubits is 5.
\item Bob applies $U_{work}^\dagger$ to the work qubit located at $I$.
\item Swap back the work qubit from location $I$, back to its original location
at $q_0$.
  
Note: due to the swap-back steps, the ancilla qubits are back into their
original locations. 
\item Alice measures the work qubit in the computation basis.
\item Measure each ancilla qubit in the computation basis.
[Optional: if $U_{uncillas}$ was applied at step 1, then before the last
step, we may rotate the ancilla qubits back,
namely apply $U_{ancillas}^\dagger$.]
\end{enumerate} 

The required threshold for quantumness is a fidelity of 2/3 for the work qubit, and also 2/3 for
each ancilla qubit. See Figure~\ref{fig:do-nothing} for the circuit,
and Figure~\ref{fig:do-nothing-histogram} in Section~\ref{sec:results}, 
for the way results are obtained,
when focusing on a single qubit.


\begin{figure}[H]
    \centering
    \includegraphics[width=1\linewidth]{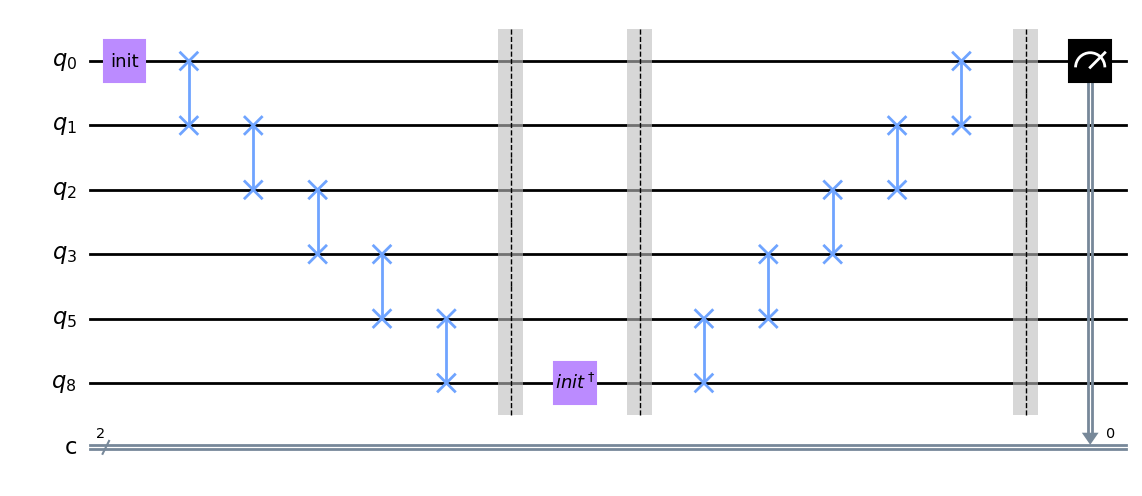}
    \caption{Do nothing protocol with $n=6$ qubits}
    \label{fig:do-nothing}
\end{figure}

A comment about the randomness expected for the results: 
In a single experiment repeated 1000 times (or 10000), randomness
is only due to the measurement of a quantum state.
However, when running the same code again, there
is also randomness due to the
randomly chosen gates (in some of our protocols). 
Additional randomness is expected as the simulated noise
of the circuit is updated by IBM (including SPAM ---
state preparation and measurement errors).

\item The super-dense coding protocol --- $[2;1]$:
\begin{enumerate}

\item Reset the two work qubits, $q_0$ and $q_1$, and the four ancilla qubits to zero.
\item Alice initializes the work qubits to a singlet state via a
transformation $U_{singlet}$.
\item Apply swap gates to move $q_1$ to Bob's node at the $I$'th
location. 

In case we choose $I=q_8$, the distance (i.e., number of swaps) is $L=4$, and
the number of relevant ancilla qubits is 4.
\item Bob applies one of the four Pauli operators to the work qubit located at $I$. The choice of which operator to apply depends on the values of the two classical bits Bob wishes to communicate to Alice. 
\item Swap the work qubit from location $I$ back to its original location
at $q_1$.  Note: due to the swap-back steps, the ancilla qubits are back into
their original locations. 
\item Alice measures the two work qubits in the Bell basis. 

\end{enumerate}


The required threshold for quantumness is a fidelity of 0.5 for the two work qubits.
See Figure~\ref{fig:superdense} for the circuit,
and Figure~\ref{fig:superdense-histogram} in the next section, 
for the way results are obtained,
when focusing on the two qubits.

\begin{figure}[H]
    \centering
    \includegraphics[width=1\linewidth]{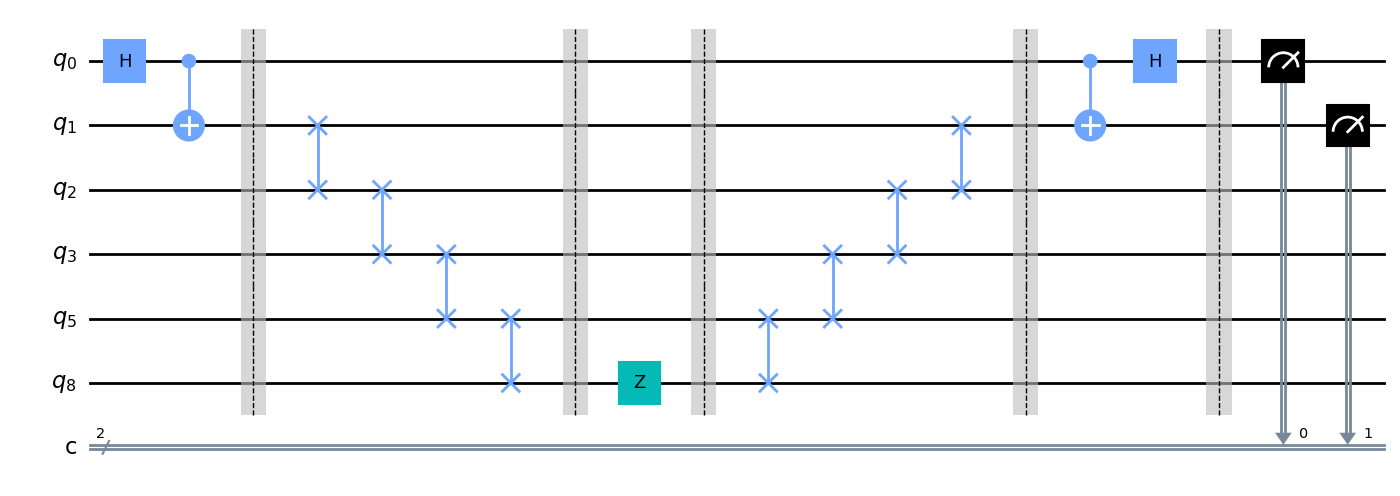}
    \caption{Superdense coding protocol with $n=6$ qubits. 
In this example, only
the Z-gate is applied.}
    \label{fig:superdense}
\end{figure}

\item The Bell-state transfer protocol --- $[2;2]$: 

\begin{enumerate}

\item Reset the two work qubits, $q_0$ and $q_1$, and the four ancilla qubits to zero.
\item Alice initializes the work qubits to any of the four 
Bell states via a
transformation $U_{Bell-state}$.
\item Apply swap gates to move $q_1$ to Bob's node at the $I$'th location.
\item Apply swap gates to move $q_0$ to Bob's node at the $I-1$'th location. 

In case we choose $I=q_8$, the distance (i.e., number of non-internal swaps)\footnote{An {\em internal swap} is a swap between qubits of the same party, e.g. Alice.} 
is $L=3$, and
the number of relevant ancilla qubits is 4.

\item Bob measures the two work qubits in the Bell basis. 

Note: in this protocol we do not apply the swap back steps, hence the ancilla
qubits are shifted by two locations upwards. 

\end{enumerate}

The required threshold for quantumness is a fidelity of 0.5 for the two work qubits.
See Figure~\ref{fig:Bell-state} for the circuit.


\begin{figure}[H]
    \centering
    \includegraphics[width=1\linewidth]{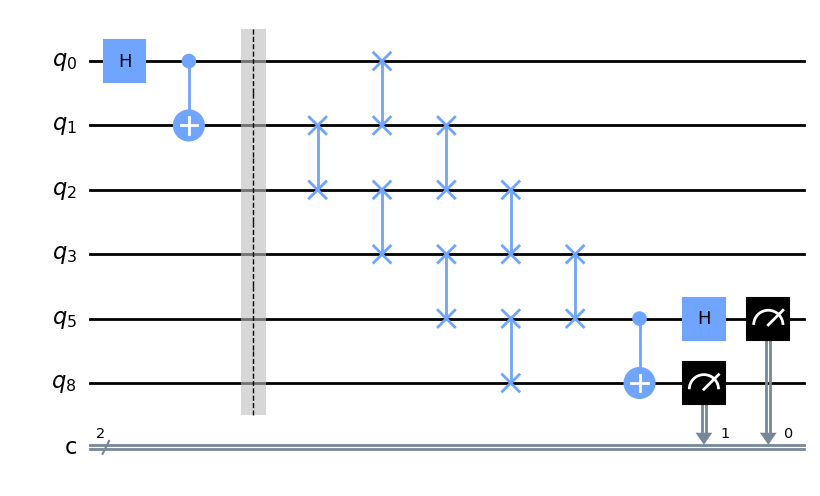}
    \caption{Bell-state transfer with $n=6$ qubits}
    \label{fig:Bell-state}
\end{figure}

\item The teleportation protocol --- $[3;1]$:

\begin{enumerate}

\item Reset the three work qubits, $q_0, q_1, q_2$ and the three ancilla qubits to zero.
\item Alice initializes $q_1$ and $q_2$ to a singlet state via
the transformation $U_{singlet}$.
\item Apply swap gates to move $q_2$ to Bob's node at the $I$'th location.

In case we choose $I=q_8$, the distance (i.e., number of swaps)
is $L=3$, and
the number of relevant ancilla qubits is 3.

\item Alice initializes $q_0$, the ``teleported qubit'', to a random state via a
transformation $U_{teleport}$.  

\item Alice measures the two work qubits, 
$q_0$ and $q_1$, in the Bell
basis. She sends the classical outcome of two bits to Bob.

\item Bob applies a Pauli operator to his work qubit, 
located at $I$, the operator is chosen according to the values of the 
two classical bits he received from Alice. As a result, $q_0$ has been
teleported to location $I$.

Note: in this protocol we do not apply the swap back steps, hence the ancilla
qubits are shifted by one location upwards. 

\item Bob applies $U_{teleport}^\dagger$ to the work qubit located at $I$.
Bob measures it to find it is in the state $\ket{0}$.
Required fidelity threshold for quantumness: 2/3 for the teleported qubit.
\end{enumerate}

The required threshold for quantumness of the teleported qubit
is 2/3.
See Figure~\ref{fig:teleportation} for the circuit.

\begin{figure}[H]
    \centering
    \includegraphics[width=1\linewidth]{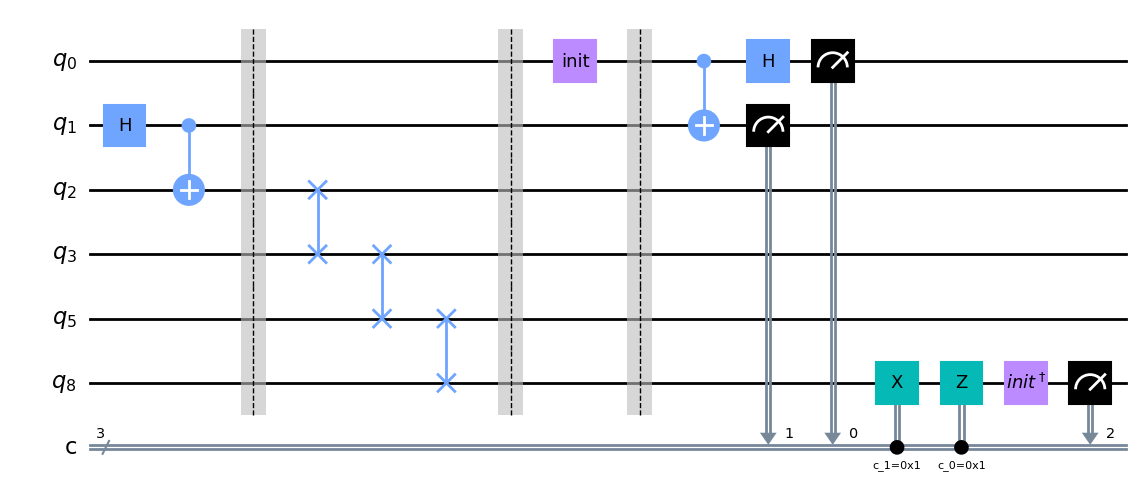}
    \caption{Teleportation with $n=6$ qubits.}
    \label{fig:teleportation}
\end{figure}

\item Entanglement swapping protocol --- $[4;2]$:

\begin{enumerate}

\item Reset the four work qubits $q_0, q_1, q_2, q_3$ and two ancilla qubits to zero.
\item Alice initializes $q_0$ and $q_1$ 
to a singlet state via
the transformation $U_{singlet}$, and she also initializes $q_2$ and $q_3$ 
to a singlet state via
the transformation $U_{singlet}$.
\item Apply swap gates to move $q_3$to Bob's node at the $I$'th location.
\item Apply swap gates to move $q_1$ is transferred 
to Bob's node at the $I-1$'th location. 

Note that the swap of $q_1$ with $q_2$ is internal at Alice's location.
Note that the swap of the original $q_3$ from location $I-1$ to location 
$I$ is internal at Bob's location.
We do not count internal swaps at Alice's site and at Bob's site,
hence for $N=6$, namely when Bob is Alice's near neighbor, $L=1$.

In our case, we choose $I=q_8$, hence 
the distance (i.e., number of non-internal swaps) 
is $L=1$, and
the number of relevant ancilla qubits is 2.

Note: there are architectures for which we may avoid the internal swaps
due to having
two disjoint paths of the same minimal length $L$, 
one path for $q_1$ and another path for
$q_3$.  In general we avoid such options, but in this work we will mention
two cases in which this option makes most sense, the ladder architecture of 
some IBM-Q computers, and the ion-trap architecture.

\item Alice measures the two work qubits she kept (now in locations $q_0$ and $q_1$) in the Bell basis. 
Note: Alice's measurement forces Bob's two qubits to become entangled,
and his entangled state is then known to her. 

\item Bob measures his qubits in the Bell basis. 
Bob is expected to have an identical Bell state outcome as the one 
Alice obtains.

Note: in this protocol we do not apply the swap back steps, hence the ancilla
qubits are shifted by two locations upwards. 

\end{enumerate}

The required threshold for quantumness is 0.5 for Bob's two work qubits. 
See Figure~\ref{fig:ent-swapping} for the circuit, and 
Figure~\ref{fig:ent-swapping-statistics} for the resulting statistics,
in Section~\ref{sec:results}.

\begin{figure}[H]
    \centering
    \includegraphics[width=1\linewidth]{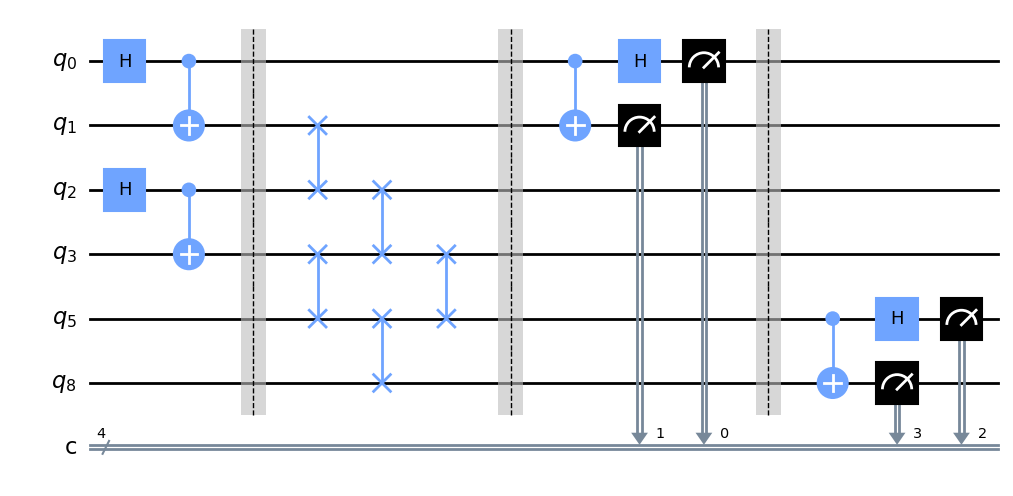}
    \caption{Entanglement swapping with $n=6$ qubits.}
    \label{fig:ent-swapping}
\end{figure}

\item  The generalized do-nothing protocol, do-nothing($\{M\}$) ---
$[M;M]$ (note the condition  $n \ge 2M$):
\begin{enumerate}

\item Reset the $M$ qubits and the $n-M$ ancilla qubits to zero.

\item 
Alice initializes $q_0 \ldots q_{M-1}$ qubits, which are 
required to be $M$ near-neighboring
work qubits in some substructure, to a random state via a transformation
$U_{work}$.  

[here we demonstrate a special partially-random case 
by the following sequence of gates: random(0) on $q_0$,
cnot($q_0 \rightarrow q_1$), random(1) on $q_1$, cnot($q_1\rightarrow q_2$), random(2) on $q_2$]
\item Apply swap gates to move $q_2$ to Bob's node at the $I$'th location, followed by swap gates to move
 $q_1$ to location $I-1$ and on $q_0$ to location $I-2$. 

In case we choose $I=q_8$, and $M=3$, 
the distance (i.e., number of non-internal swaps) is $L=1$, and
the number of relevant ancilla qubits is 3.

\item Bob applies $U_{work}^\dagger$ to the work qubit located at $I-1$,
$I-1$, and $I$.
.
\item Swap back the work qubits their original locations.
  
Note: due to the swap-back steps, the ancilla qubits are back into their
original locations. 
\item Alice measures the work qubits in the computation basis.

Note that the regular do-nothing is identical to do-nothing($\{1\}$).
\end{enumerate}

The required threshold for quantumness is 2/3 for each returned qubit. 
We implemented the case of $M=3$, $L=1$. 
See Figure~\ref{fig:generalized-do-nothing} for the circuit. 
\begin{figure}[H]
    \centering
    \includegraphics[width=1\linewidth]{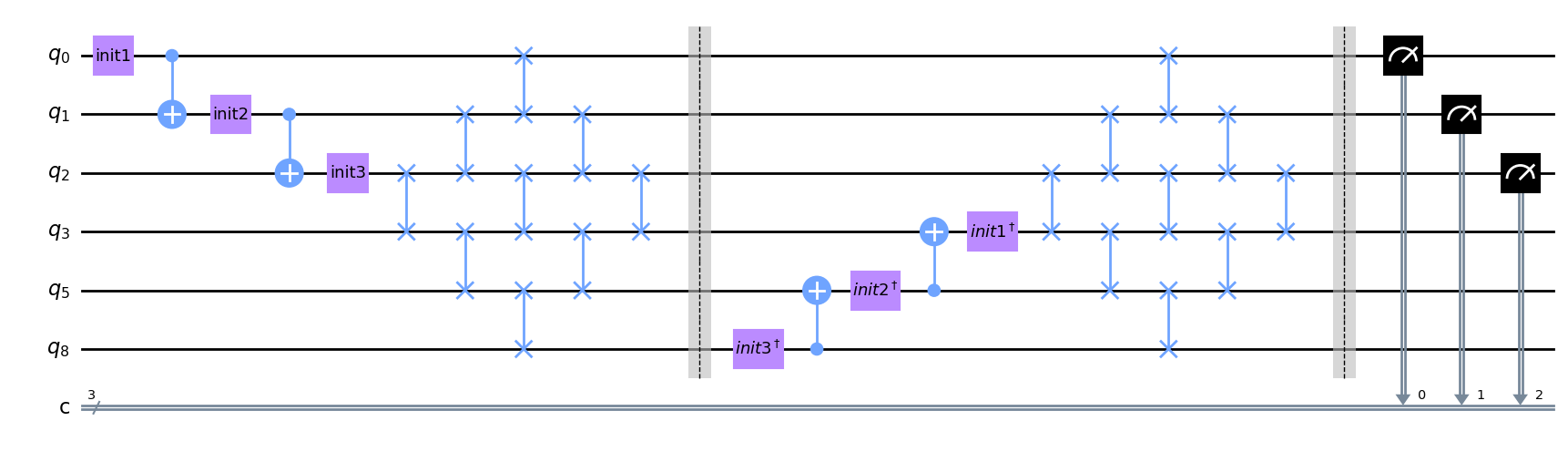}
    \caption{Generalized do-nothing $\{M=3\}$ with $n=6$ qubits.}
    \label{fig:generalized-do-nothing}
\end{figure}

\item The cat-state $\{M;J\}$ protocol --- 
 $[M;J]$ (note the conditions  $n \ge M+J$, and $J \ge2$):

\begin{enumerate}

\item Reset the $M$ qubits and the $n-M$ ancilla qubits to zero.

\item 
Alice initializes $q_0 \ldots q_{M-1}$ qubits, which are 
required to be $M$ near-neighboring work qubits 
in some substructure, to a cat state, 
$\ket{00\ldots 0} + \ket{11\ldots 1}$ 
via a transformation
$U_{cat}$.  
  
\item Apply swap gates to qubit $q_{M-1}$ to Bob's node at the $I$'th location. 
Then swap $J-1$ more near-neighbor work qubits, via the same path, 
to Bob so that the entire substructure of the
near-neighboring  $J$ qubits is reconstructed at Bob's site.

We explicitly present here three different examples, 
while we satisfy the requirement that there is no overlap
between Alice's location and Bob's location:

Note that the case of $\{M=2;J=2\}$ is identical to the Bell state
transfer protocol, hence it is not listed as an option below.

Option 1:
Let $M=3$ and $J=2$: apply swap gates to move $q_2$ to
Bob's node at the $I$'th  location, followed by swap gates to move
$q_1$ to location $I-1$. 

Qubit $q_0$ is measured by Alice in the Hadamard
basis in the next step of the protocol.

Option 2:
Let $M=3$ and $J=3$: apply swap gates to move $q_2$ to
Bob's node at the $I$'th  location, followed by swap gates to move
$q_1$ to location $I-1$ and on $q_0$ to location $I-2$. 

Option 3:
Let $M=4$ and $J=2$: apply swap gates to move $q_3$ to Bob's node at the $I$'th  location, followed by swap gates applied
on $q_2$ to location $I-1$.

Qubits $q_0$ and $q_1$ are measured by Alice in the Hadamard
basis in the next step of the protocol.

In case we choose $I=q_8$ 
the distance (i.e., number of non-internal swaps) is $L=2$ in the first 
option, and $L=1$ in the last two examples since Alice and Bob are then
near neighbors.

\item Alice measures her $M-J$ qubits in the Hadamard basis. Bob keeps two
near-neighboring qubits un-measured and for $J>2$, 
he measures all the remaining
$J-2$ qubits in the Hadamard basis. 

\item Alice informs Bob of the results in $\{+,-\}$ due to measuring in the Hadamard basis. 

If the total number of minuses, for Alice and Bob together, is even Bob does nothing to the
remaining pair of qubits. 

If the total number of minuses is odd Bob applies $\sigma_z$.

The resulting state of the two qubits at Bob's hands is always 
$\ket{\Phi_+}$ for a noiseless protocol.

\item  Bob conducts a Bell measurement.

Note: in this protocol we do not apply the swap back steps, 
hence the ancilla
qubits are shifted by $J$ locations upwards. 

\end{enumerate}

The required fidelity threshold for quantumness is 0.5 on Bob's two remaining qubits. 
We implemented option 3 --- the case of $\{M=4;J=2\}$.
See Figure~\ref{fig:cat-state}

\begin{figure}[H]
    \centering
    \includegraphics[width=1\linewidth]{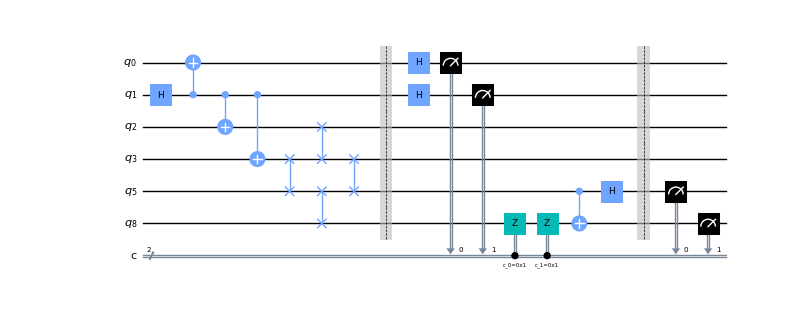}
    \caption{Cat-state protocol with $n=6$ qubits, and with 
$[M=4;J=2]$}
    \label{fig:cat-state}
\end{figure}
\end{enumerate}


From a protocol, and successful vs.\@ failed paths, to effective number of qubits:
We use two methods to define a success of a chip or a substructure of a chip (subchip) so
to claim it has $K$ effective qubits for a protocol.
One method, optimal for protocols 1--5 is to find a subchip such that
for any relevant Alice and Bob nodes the protocol succeeds. 
As a simple example, if we run the {\em do-nothing} protocol successfully on any path for the $K$-qubit
subchip, yet fail on one or more paths when we add another qubit to obtain a
$K+1$ substructure, we conclude that the $N$-qubit chip has a $K$-qubit
subchip. In other words, there are $K$ effective qubits for the do-nothing
protocol on that subchip.
Another method, optimal for protocols 6--7 is to find $M$ or to find $M$ and $J$
such that the protocol still works for a specific distance, while it stops
working if we increase $M$ or $J$. 

Noise model: It serves no purpose to use a noise-free model, as even $L = \infty$ will succeed.
One can either work with a less realistic simulator, naively guessing the noise
parameters which are unavailable in the public domain,
or work with official simulators of companies that supply sufficiently easy access to them. 
Working with the actual quantum devices is usually much slower due to long queues.

\subsection{A fully checked simulated subchip of Kolkata - a trivial example}

If we define the linear subset containing the six qubits from
$q_0$ to $q_8$ as the subchip we wish to analyze,
clearly it is a shortest linear subset as there is no shorter path
between $q_0$ and $q_8$, and also any subset of it, is still
a shortest linear subset. 

Then if for any of the seven protocols one would have checked all
possible options, one would say that the entire subchip is checked.
If furthermore for all its subsets the protocol paths the threshold,
this means that the subchip is proven to show ``quantumness''.
The worst result in such a case would be the number provided to the protocol
vector. 
For example, for the basic 
do-nothing protocol there are 30 shortest linear subsets to check
(15 with increasing qubit numbers and 15 with decreasing qubit numbers),
and the worst fidelity result among all those, is the one provided
to the protocol vector. If it above the quantumness threshold, quantmness
is satified for that protocol on that subchip.

The protocol vector would include such results for all five basic
protocols. If for all five protocols quantumness is satisfied to all relevant 
distances, the subchip passes the quantumness test.
Following previous explanations, the threshold for the five basic 
protocols is that the vector will be above [2/3, 1/2, 1/2, 2/3, 1/2].

In addition, the two generalized protocols can also be tested, 
and in our very short string --- for all
the options --- all possible relevant linear subsets:
For the generalized do-nothing, if we ignore the 30 tests mentioned above,
one needs to check all options for tranfering 2 adjacent qubits 
(12 options for that), or three (just two options for that).
For the cat-state protocol, if we ignore the trivial case
which is the Bell-state transfer, one needs to check all options
for 
$\{M=3,J=2\}$ (6 options), 
$\{M=3,J=3\}$ (two options), and  
$\{M=4,J=2\}$ (two options).

\section{Results} \label{sec:results}

\begin{figure}[H]
    \centering
    \includegraphics[width=0.5\linewidth]{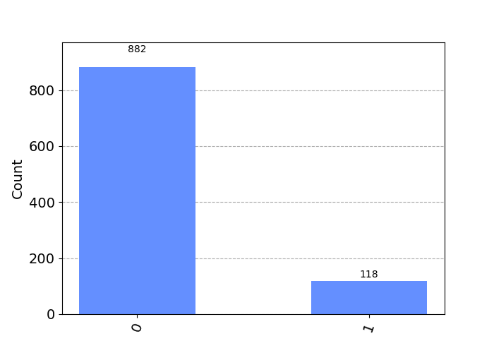}
    \caption{Result distribution for do-nothing 
when running the same gates 1000 times 
yielding a statistical fidelity of 0.882, well above the threshold of
2/3.}
    \label{fig:do-nothing-histogram}
\end{figure}

\begin{figure}[H]
    \centering
    \includegraphics[width=0.5\linewidth]{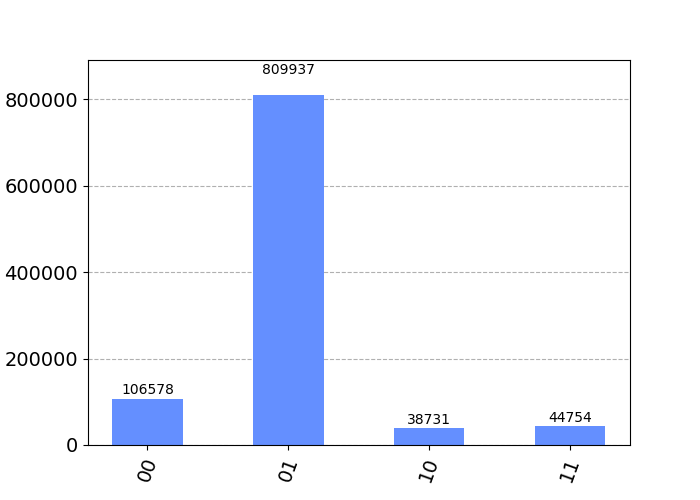}
    \caption{Result distribution for superdense coding
when running the same gates 1,000,000 times,  
yielding a statistical fidelity of 0.809, well above the threshold of
0.5.}
    \label{fig:superdense-histogram}
\end{figure}

\begin{figure}[H]
    \centering
    \includegraphics[width=1.0\linewidth]{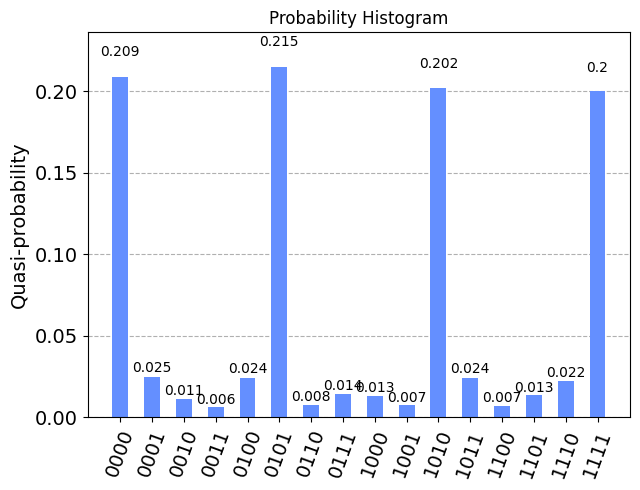}
    \caption{Entanglement swapping statistics, summing all cases of identical
Bell-states for Alice and for Bob, namely the cases 0000, 0101, 1010, and 1111,
yields the fidelity 0.826, well above the desired threshold of 0.5.}
    \label{fig:ent-swapping-statistics}
\end{figure}

In Section~\ref{sec:protocols} we explained and demonstrated as circuits 
the seven six-qubit protocols on fake IBM-Q's Kolkata. Here are the results:
First, we provide three histograms to clarify how fidelity is acquired,
see Figures~\ref{fig:do-nothing-histogram}, \ref{fig:superdense-histogram}
    and~\ref{fig:ent-swapping-statistics}, 
for the do-thing, superdense-coding and entanglement-swapping respectively. 
We then also provide results 
for the remaining protocols 
demonstrated in Section~\ref{sec:protocols}.  

In addition to the three results shown in 
the three figures, here are
the results for the two remaining basic protocols: 
For teleportation we checked the statistics by running 1000
times versus running 10,000 times. The results were 0.923 and 0.919
respectively, sufficiently near each other (a 0.5\% difference) 
and far above the threshold of 2/3. 
For Bell-state transfer the result was 0.804, well above the 0.5
threshold.

Overall, the results were above the threshold of 
[2/3, 1/2, 1/2, 2/3, 1/2], however in this demonstration only maximal 
distance on this short string was verified, so it says nothing definite
about the quantumness of this simulated subchip.

We implemented the generalized do-nothing in case of $M=3$, $L=1$. 
The results are acquired for each of the three qubits separately,
and are borderline quantum:
0.753, 0.687, and 0.704; These results are still showing quantumness, however
rather close to the threshold of 2/3 on $q_1$.
The result we obtained for cat-state $\{M=4;J=2\}$ was 0.846, 
high above the 0.5
threshold.

In contrast to the 6-qubit results presented above, in
the rest of this section we present results
that search and often 
find the boundary between classicality and quantumness. Furthermore,
in some cases we make use of the classicality results to remove some 
of the qubits and define an effective subchip which is quantum for
the relevant protocol, hence we also find the number of effective qubits.

\subsection{IBM-Q Melbourne}~\label{ssec:res-Melbourne} 

\begin{figure}[H]
    \centering
    \includegraphics[width=0.5\linewidth]{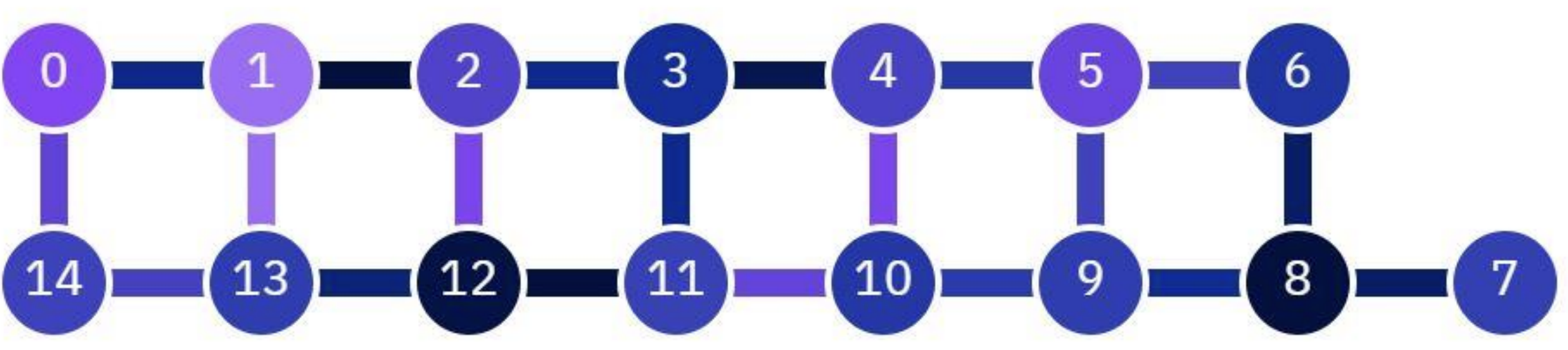}
    \caption{IBM's Melbourne.}
    \label{fig:Melbourne-chip}
\end{figure}

We tested the presented five basic protocols on a simulation of one of IBM's devices called \emph{ibm\_Melbourne}, 
and acquired the relevant device protocols vector and it's effective qubits.
    See Figure~\ref{fig:Melbourne-chip} for the device connectivity map.\footnote{At a later time, IBM disconnected a qubit that was badly functioning, leaving the device with only 14 qubits.}
    The simulation was done using the Qiskit software package, and the noise model parameters were based on the characterization parameters of the device done by IBM and saved as a noise model
    named Fake-Melbourne. The noise model consists of one- and two-qubit gate errors in the form of depolarizing error channel, energy relaxation errors, dephasing errors
    and non-correlated measurement errors~\cite{qiskit_aer_noise_tutorial}. 
    
\begin{figure}[H]
    \centering
    \includegraphics[width=0.9\linewidth]{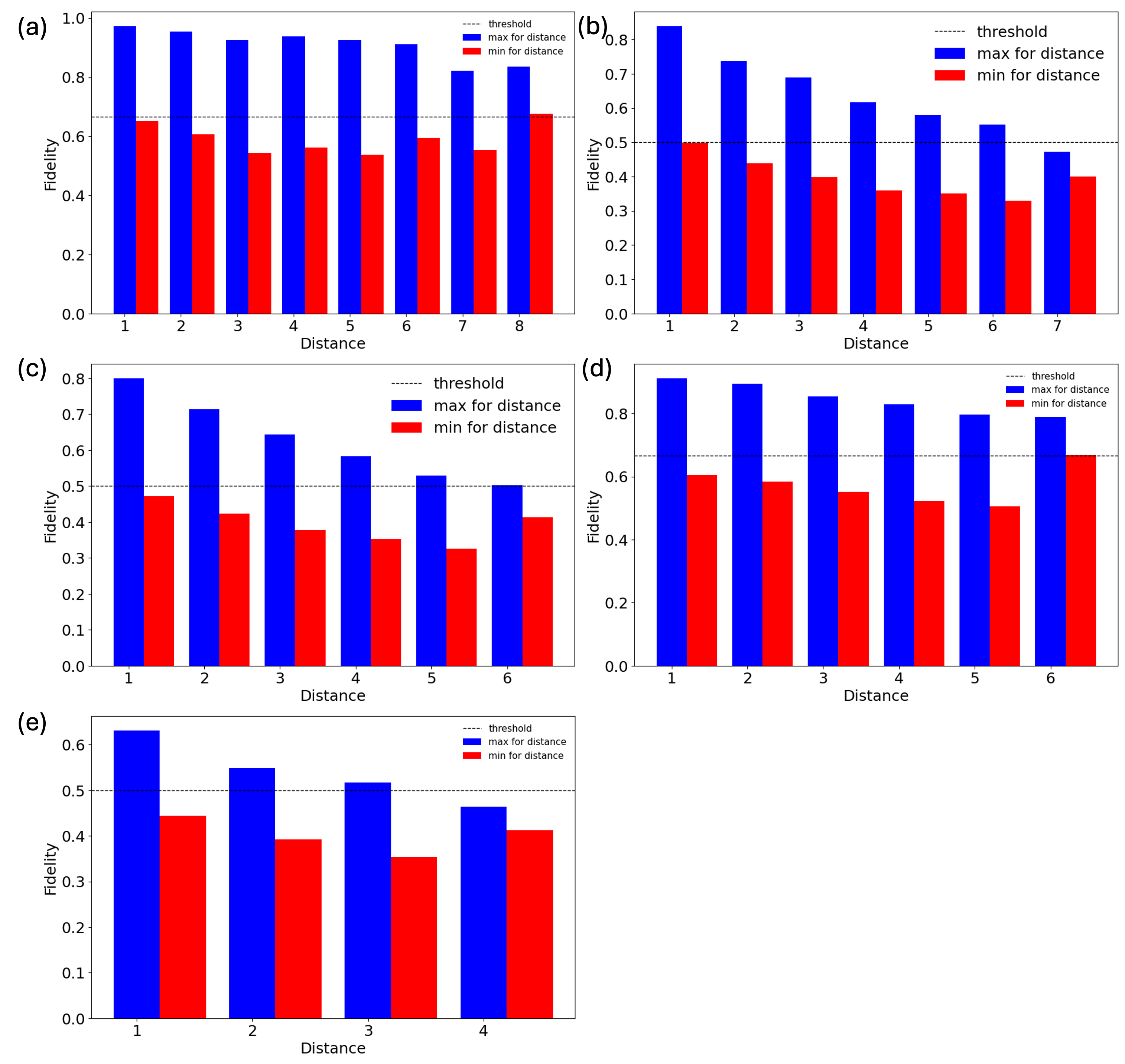}
    \caption{IBM's Melbourne simulation results on the different protocols, run
on all possible shortest linear subsets in the device. 
(a) results of the do-nothing protocol. (b) results of superdense-coding protocol. (c) results of the Bell-state transfer protocol. (d) results of the teleportation protocol. (e) results of the entanglement swapping protocol.}
    \label{fig:Melbourne-simulation-results}
\end{figure}

    Each protocol was run on each possible shortest linear subset  
between the device's qubits (where only shortest paths between qubits were taken into account),
    while for each path we used 10000 shots to obtain the fidelity of the protocol. Using the protocols' fidelities, we found, for each protocol, the maximal distance
    which passes the protocols' threshold. Meaning that there is at least one
set of qubits on which we can run the relevant protocol to this maximal distance
and keep their quantum properties. In addition, for each protocol, we found the
subchip comprised of all the paths on which we can run the protocol and pass the
threshold. The minimal and maximal fidelities of each protocol run on all of the
possible shortest linear subsets are presented in Figure~\ref{fig:Melbourne-simulation-results}. 
     In this diagram and similar ones, we present per specific length, the best case (the path with maximal fidelity). Namely, there exist a path of that length with this fidelity, in blue. We also present per specific length, the worst case (with minimal fidelity). Namely, all paths of that length have at least this fidelity, shown in red. Having a blue result above the threshold means we have quantumness for at least one path of that length, and a red result above the threshold means we have quantumness for all paths of that length. 
  The resulting protocols vector of the whole device is 
[0.5373, 0.3303, 0.3261, 0.5059, 0.3543], 
where each component is the minimal fidelity achieved for the corresponding 
protocol. 

The fidelity obtained for the longest distance is higher 
than some of the shorter distances for some of the protocols. 
The cause is a very large readout error of qubit 6.  
Because of its location in the connectivity map, qubit 6 is an ancilla qubit in
the longest distance paths of the device for most of the proposed protocols, and
therefore is not being measured in these paths, and does not then disturb the 
fidelity. It is important to clarify that it is a problem of the specific
qubit, and hence we would have recommended to IBM, if the real chip  
was still available, to remove that qubit, similarly to their 
removal other bad qubits when needed.

Compared to the desired protocol vector where each number should be
above [2/3, 1/2, 1/2, 2/3, 1/2] the result is not good, and the chip cannot be
considered to show quantumness in resepct to any of the protocols.

\begin{figure}[H]
    \centering
    \includegraphics[width=0.9\linewidth]{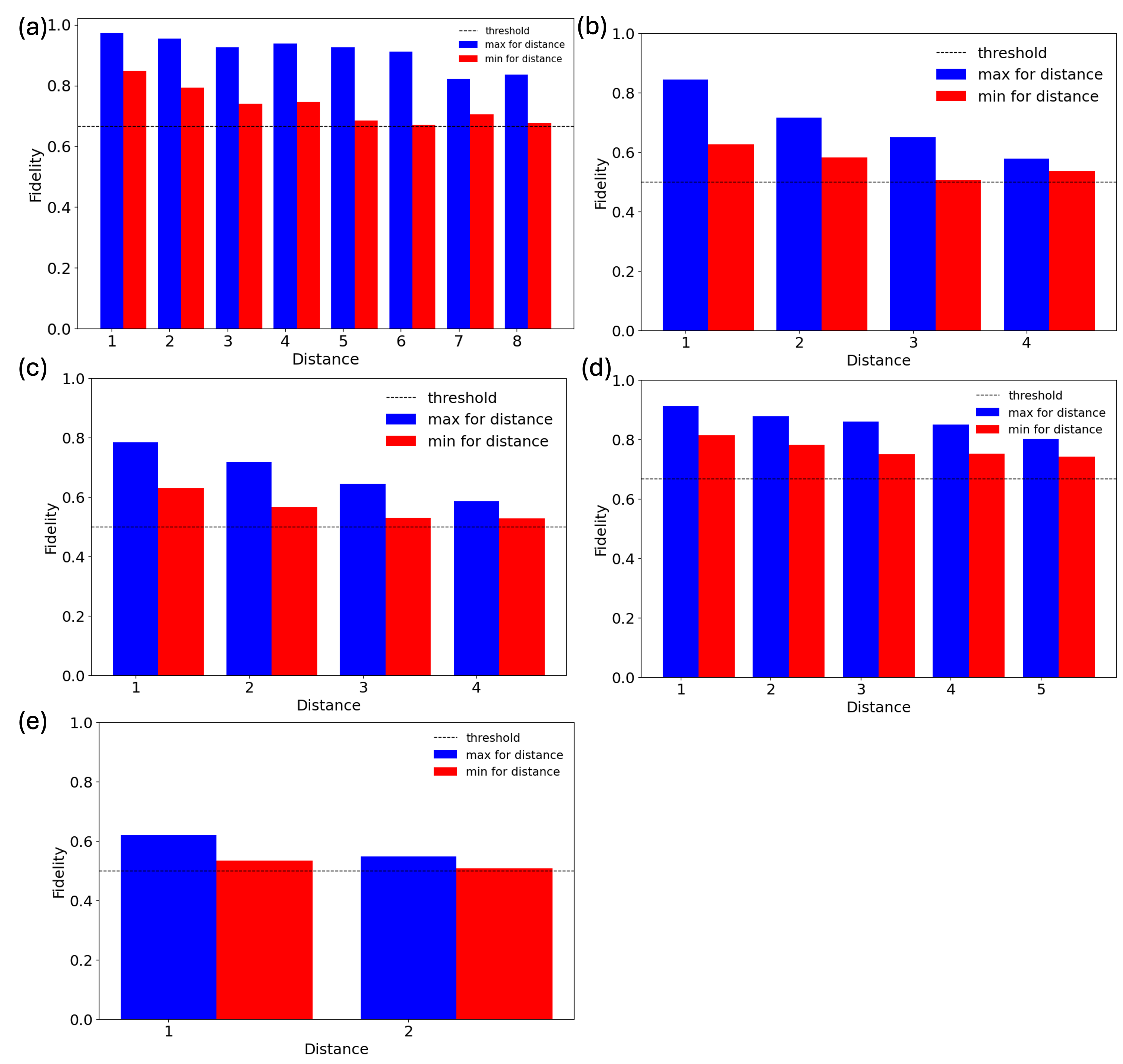}
    \caption{IBM's Melbourne simulation results on the different protocols, run
on all possible shortest 
linear subsets after excluding some of the qubits in the device. (a) results of the do-nothing protocol after excluding qubit 6. (b) results of superdense-coding protocol after excluding qubits $q_6\dash q_9$, $q_{13}$ and $q_{14}$. (c) results of the Bell-state transfer protocol after excluding qubits $q_6\dash q_8$ and $q_{14}$. (d) results of the teleportation protocol without excluding any qubits. (e) results of the entanglement swapping protocol after excluding qubits $q_5\dash q_9$ and $q_{13}$.}
    \label{fig:Melbourne-simulation-results-filtered}
\end{figure}

For each protocol, we examined the results, and excluded some qubits in order to achieve a subchip where the minimal fidelity for all of the distances of the protocol passes the threshold, meaning that for every possible path between a set of qubits within this subchip, the protocol maintains it's quantumness, while trying to keep as much qubits as possible for each protocol. The minimal and maximal fidelities of each protocol run on all of the 
possible shortest linear subsets in the remaining subchips are presented in
Figure~\ref{fig:Melbourne-simulation-results-filtered}. The protocols vector
achieved in this way is [0.6692, 0.5055, 0.5272, 0.7413, 0.5066], 
slightly above the desired thresholds which are  
[2/3, 1/2, 1/2, 2/3, 1/2].

\begin{figure}[H]
    \centering
    \includegraphics[width=0.9\linewidth]{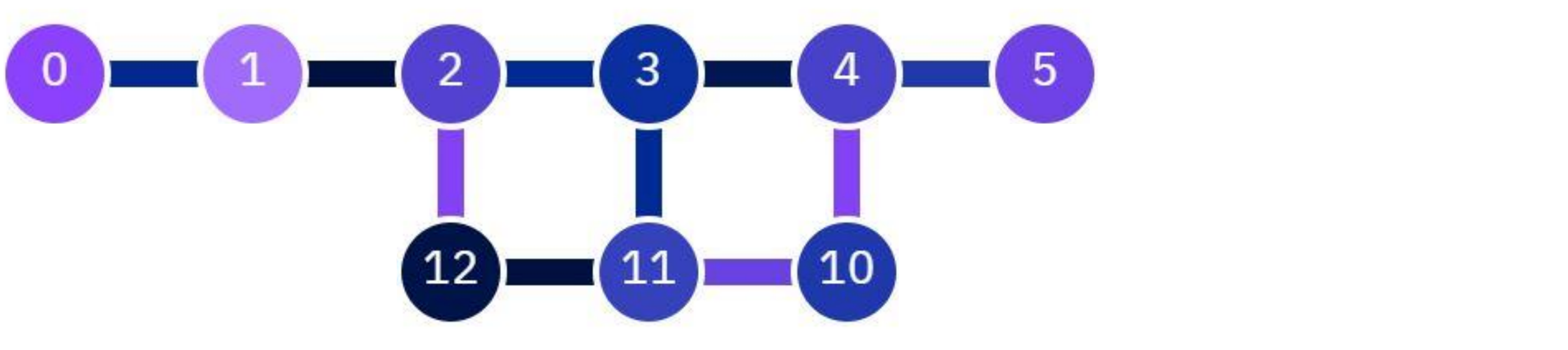}
    \caption{IBM's Melbourne common subchip for all the basic protocols}
    \label{fig:Melbourne-common-subchip}
\end{figure}

\begin{figure}[H]
    \centering
    \includegraphics[width=0.9\linewidth]{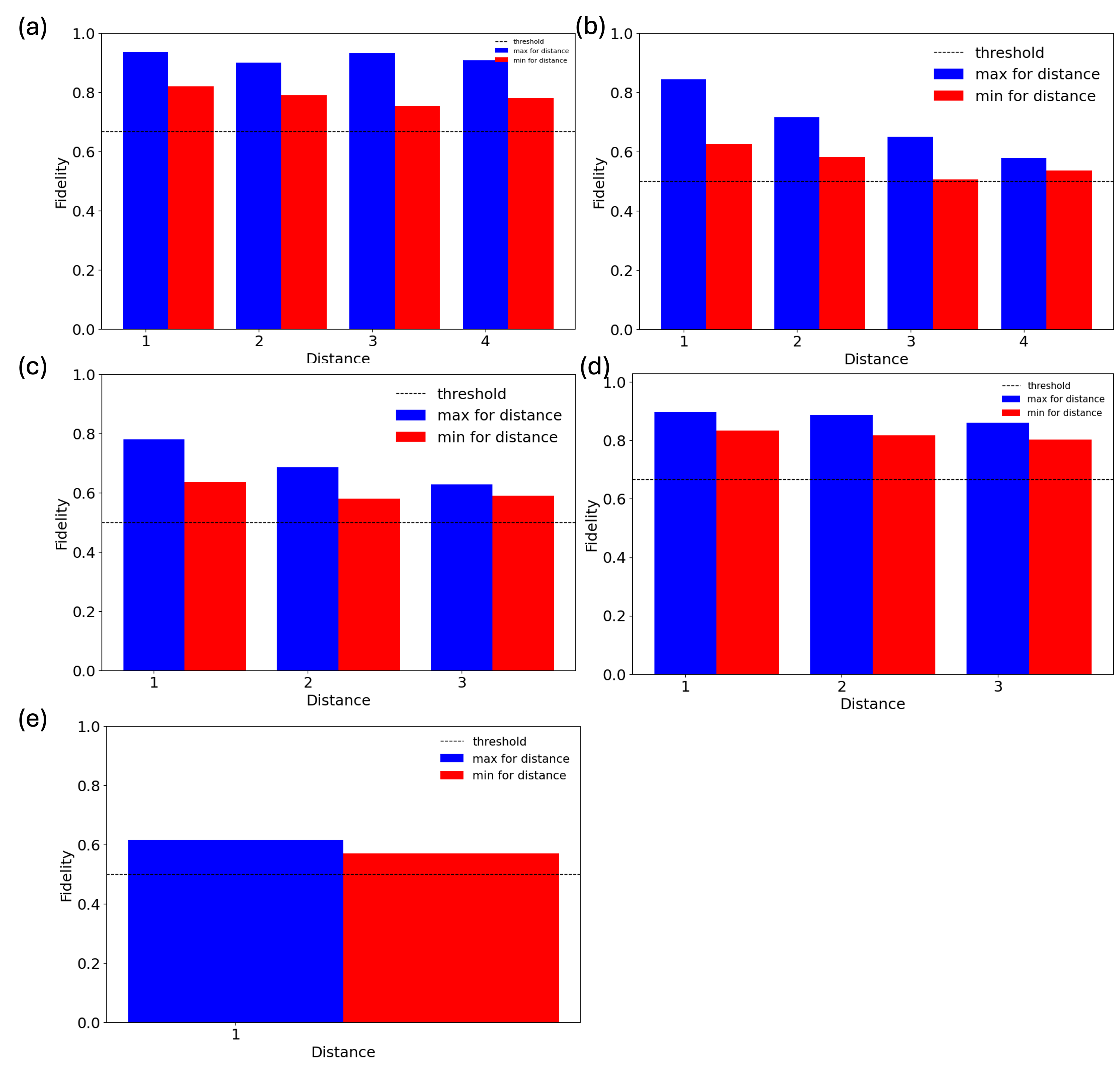}
    \caption{IBM's Melbourne simulation results on the different protocols, run
on all possible shortest 
linear subsets after excluding qubits $q_6\dash q_9$, $q_{13}$ and $q_{14}$. (a) results of the do-nothing protocol. (b) results of superdense-coding protocol. (c) results of the Bell-state transfer protocol. (d) results of the teleportation protocol. (e) results of the entanglement swapping protocol.}
    \label{fig:Melbourne-simulation-results-sub-chip-filtered}
\end{figure}

The common subchip that passes the threshold for each one of the basic protocols
for all of the shortest 
linear subsets within it, consists of qubits $q_0\dash q_5$, and $q_{10}\dash q_{12}$. See Figure~\ref{fig:Melbourne-common-subchip} for its connectivity map. These qubits form the common effective subchip for the five basic protocols. We found the protocols vector for this subchip to be [0.7539, 0.5055, 0.5786, 0.8034, 0.5686]. The minimal and maximal fidelities of each protocol in this subchip are presented in Figure~\ref{fig:Melbourne-simulation-results-sub-chip-filtered}.

\begin{figure}[H]
    \centering
    \includegraphics[width=0.9\linewidth]{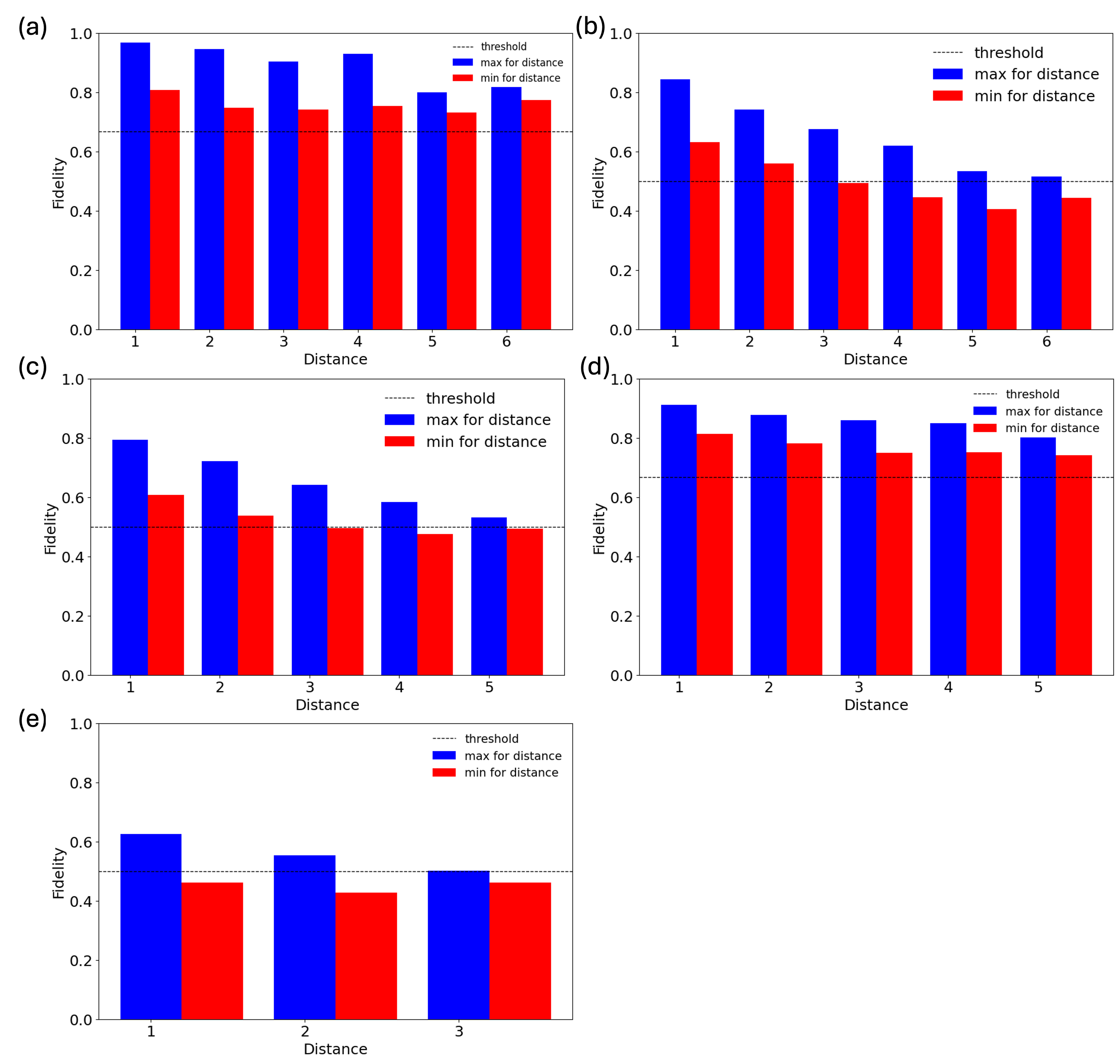}
    \caption{IBM's Melbourne simulation results on the different protocols, run
on all possible shortest
linear subsets after excluding $q_6$ and $q_7$ from the device. (a) results of the do-nothing protocol. (b) results of superdense-coding protocol. (c) results of the Bell-state transfer protocol. (d) results of the teleportation protocol. (e) results of the entanglement swapping protocol.}
    \label{fig:Melbourne-simulation-results-sub-chip-6-7-filtered}
\end{figure}

By sorting the protocols' fidelity results, and noticing which qubits are common to the worst results, we identified $q_6$ and $q_7$ as the most noisy qubits. By excluding only these qubits, one can achieve better fidelities for the different protocols, while maintaining most of the device qubits. The protocols vector of the device after excluding only these qubits is [0.7322, 0.4061, 0.4760, 0.7413, 0.4284]. The minimal and maximal fidelities of each protocol in the resulting subchip are presented in Figure~\ref{fig:Melbourne-simulation-results-sub-chip-6-7-filtered}.

\begin{figure}[H]
    \centering
    \includegraphics[width=0.9\linewidth]{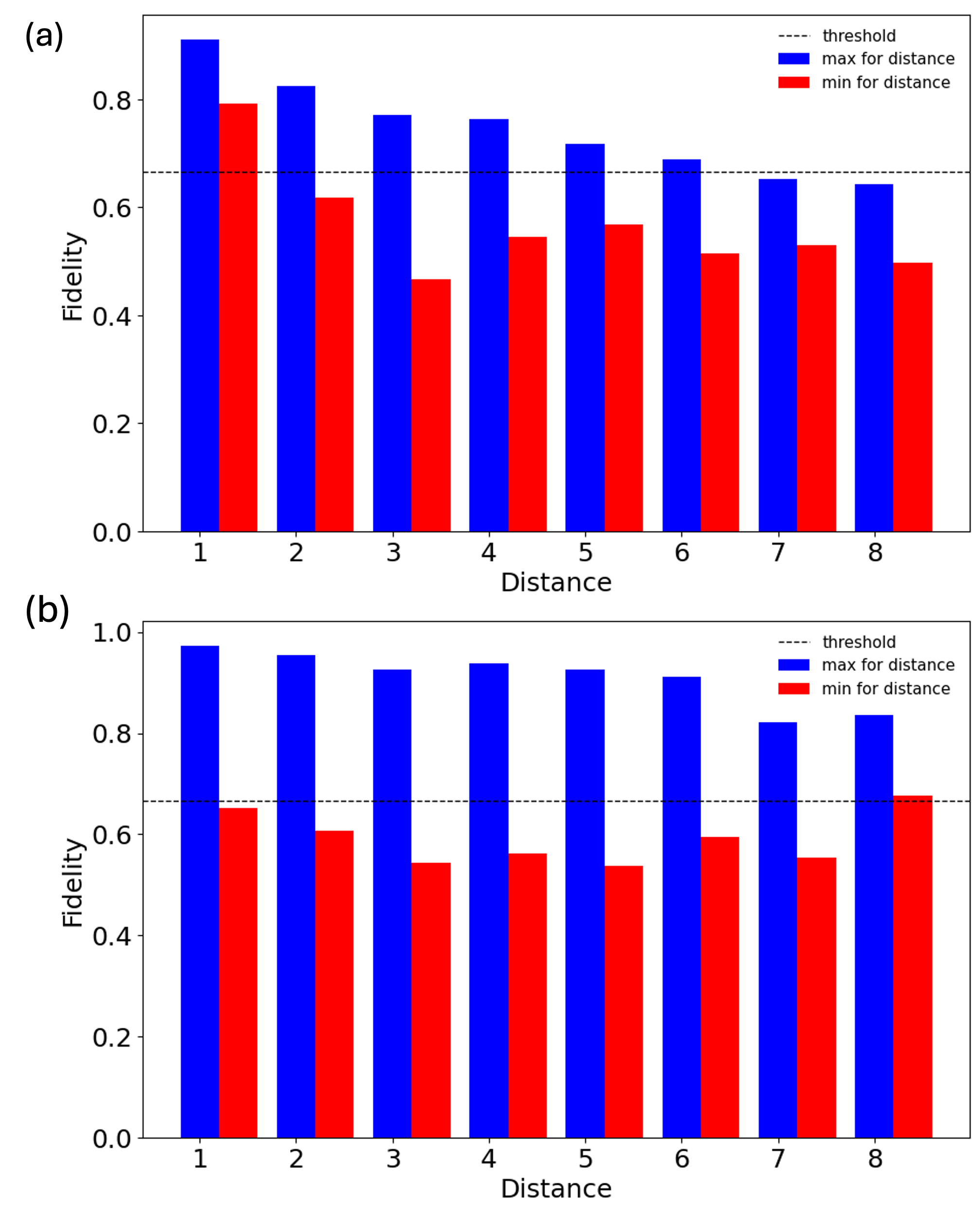}
    \caption{IBM's Melbourne real device results vs the simulation results on
the do-nothing protocol on all possible shortest linear subsets. 
(a) results of the real device. (b) the simulation results.}
    \label{fig:Melbourne-real-vs-simulation}
\end{figure}

In addition, we ran the do-nothing protocol on all the possible 
shortest linear subsets on the real ibm\_melbourne device and compared it's results to the simulation results. For each path, we used 8192 shots to acquire the fidelity of the protocol. See the results of the real device, next to the simulation results in Figure~\ref{fig:Melbourne-real-vs-simulation}. It can be seen, that there are additional errors on the real device which are not simulated, therefore the simulation fidelities are usually higher. On the real device, non of the paths at a distance larger than 6 achieved quantumness for this protocol.

\subsection{IBM-Q Kolkata}~\label{ssec:res-Kolkata} 

In this section we present our results on IBM's simulator of Kolkata, and
we also compare to one result on the real chip.

We first checked two of our basic protocols: the do-nothing protocol 
up to the maximal 
distance of $L=12$, and teleportation up to the maximally possible 
distance of $L=10$. We always included $q_0$, so we sent
$q_0$ in the do-nothing protocol, and 
teleported $q_0$ to either $q_2$ or $q_4$ (sent 
to the maximal distance beforehand, see Figure~\ref{fig:Kolkata}) 
in the teleportation protocol. 

The results showed quantumness to any distance.  
In particular, do-nothing $q_0\rightarrow q_{26} \rightarrow q_0$ yielded:
0.858 for the path via $q_2$, 0.781 for the path via $q_{13}$, and 0.801
for the path via $q_{15}$.
And in particular, teleportation of $q_2$: $q_2\rightarrow q_{26}$ and
$q_2 \rightarrow q_{24}$ (via $q_{11}$ and $q_{22}$) showed good quantumness e.g.,
to $q_{24}$ yielded 0.784.

Our quantumness results of the first experiment
suggested that Fake-Kolkata is an effective 27-qubit 
simulator for both do-nothing and teleportation. 

\begin{figure}[H]
    \centering
    \includegraphics[width=0.6\linewidth]{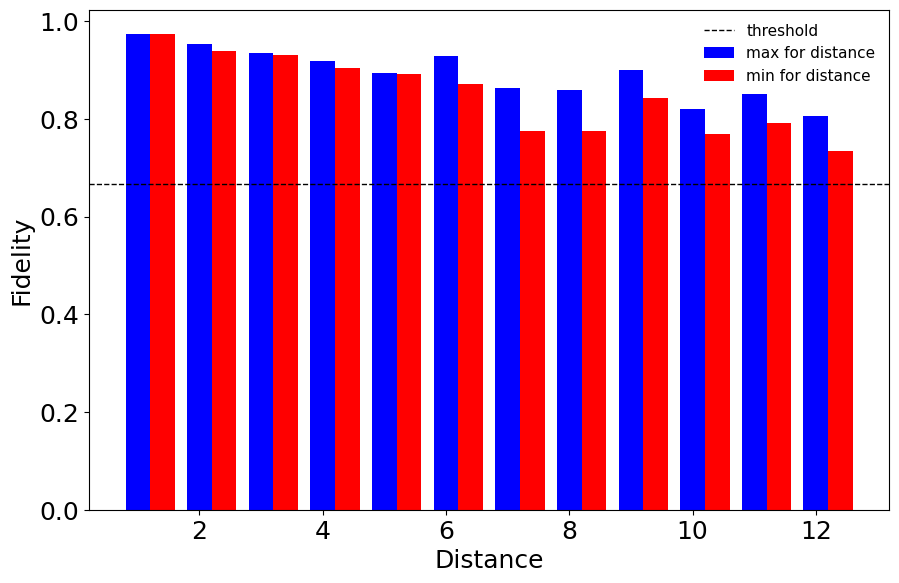}
    \caption{\textit{ibm\_kolkata} simulation results on do-nothing protocol, only paths starting from $q_0$.}
    \label{fig:Kolkata_do_nothing_simulation}
\end{figure}

\begin{figure}[H]
    \centering
    \includegraphics[width=0.6\linewidth]{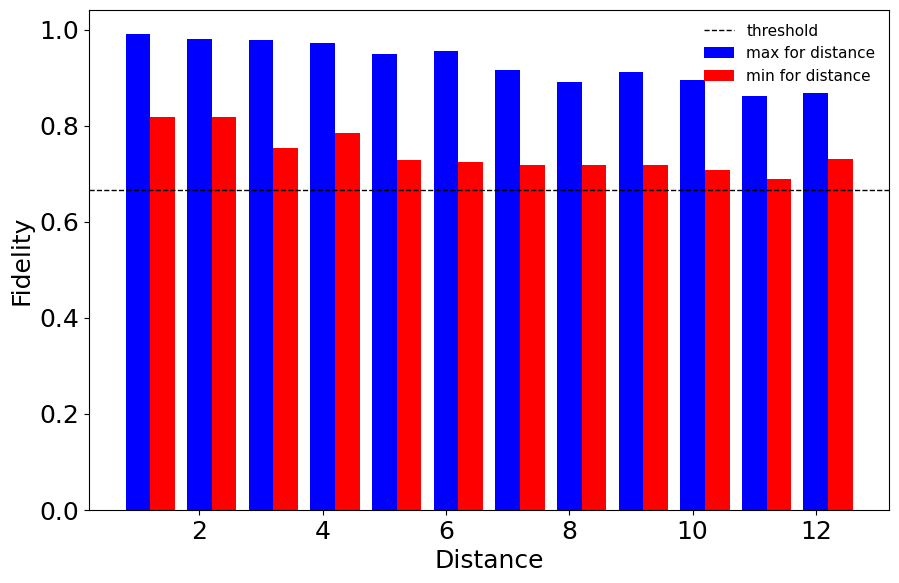}
    \caption{\textit{ibm\_kolkata} simulation results on do-nothing protocol, running on every possible shortest path (only minimal paths between 2 qubits were considered).}
    \label{fig:Kolkata_do_nothing_simulation_all_paths}
\end{figure}

\begin{figure}[H]
    \centering
    \includegraphics[width=0.6\linewidth]{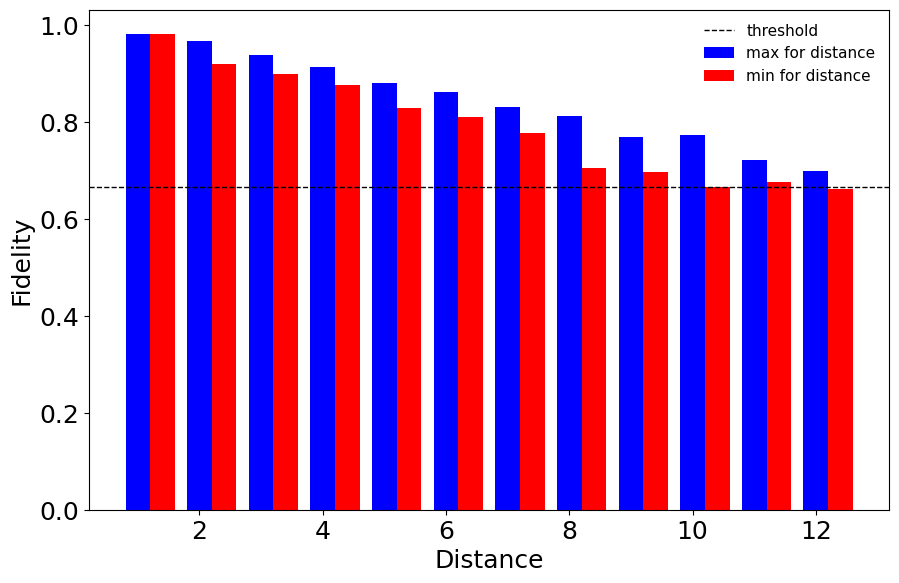}
    \caption{\textit{ibm\_kolkata} real device results on do-nothing protocol, only paths starting from qubit $0$.}
    \label{fig:Kolkata_do_nothing_real_device}
\end{figure}

In our second experiment, we applied the do-nothing protocol over all the possible shortest paths between qubits.
The results are shown in Figures~\ref{fig:Kolkata_do_nothing_simulation_all_paths}. All of the paths maintained quantumness, therefore, as the results of the first experiment suggested, all of the device's qubits form the effective qubits for this protocol.
The results show better fidelity compared to the Melbourne device's result. The hardware of this device is a newer generation of IBM's hardware, showing improvement in noise reduction.
 We also compared between the simulation results and results from the real device. For this comparison, we acquired the fidelities only for paths starting from qubit $q_0$. The results are shown in Figures~\ref{fig:Kolkata_do_nothing_simulation},~\ref{fig:Kolkata_do_nothing_real_device}. The real device's results have slightly more noise, as only some of the possible noise sources are taken into account in the simulation, given by Qiskit~\cite{qiskit_aer_noise_tutorial}.

\subsection{Beyond the linear subset on fake Kolkata}~\label{ssec:Cat-Kolkata} 

In our third Kolkata experiment, we demonstrate 
the use of the generalized protocols --- 
We chose the cat state protocol,  
and we go beyond the linear subset, in this
demonstration.  

\begin{figure}[H]
    \centering
    \includegraphics[width=0.75\linewidth]{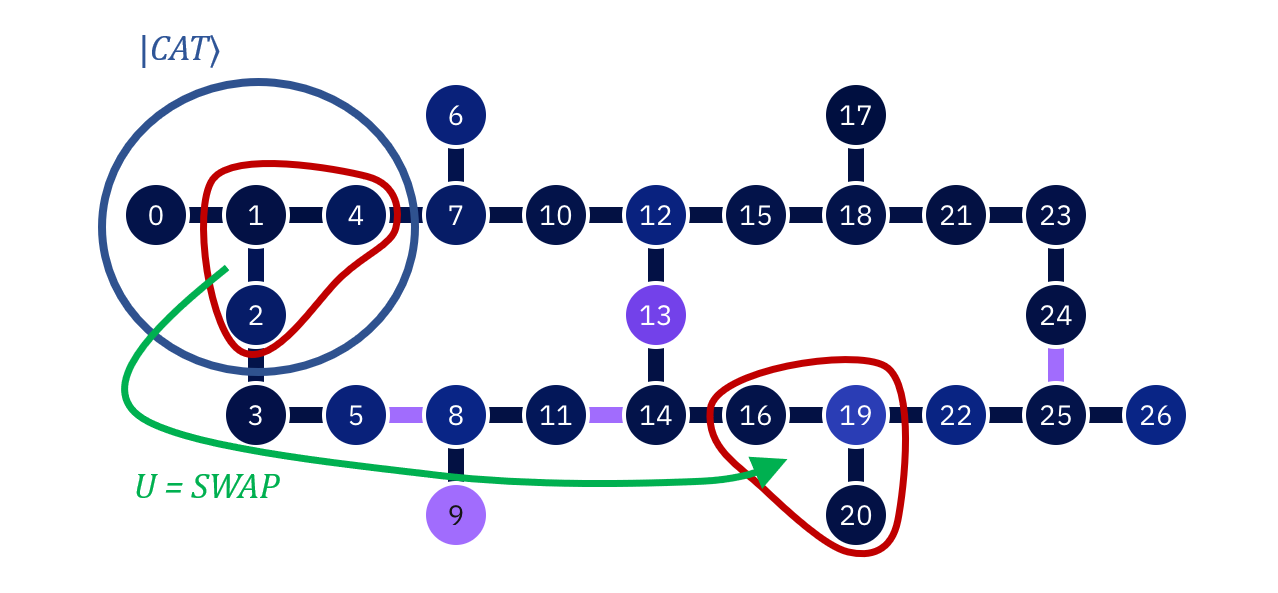}
    \includegraphics[width=0.75\linewidth]{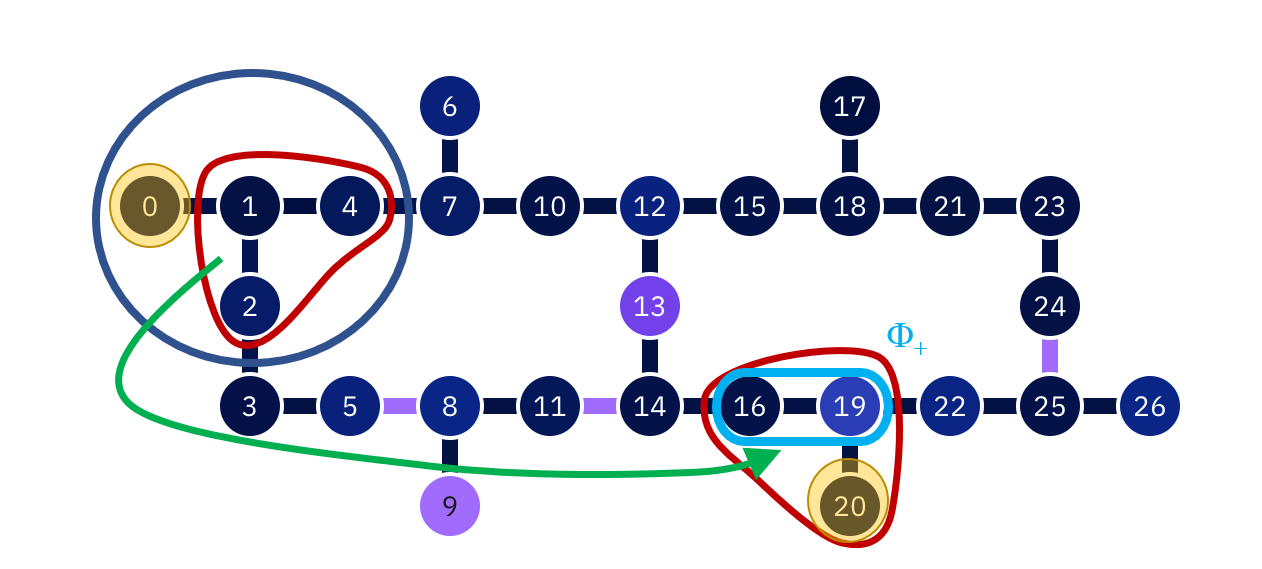}
    \caption{Failed-to-show quantumness --- the cat-state protocol,
for $M=4$ and $J=3$.}
    \label{fig:failed-cat-state-protocol}
\end{figure}

\begin{figure}[H]
    \centering
    \includegraphics[width=0.75\linewidth]{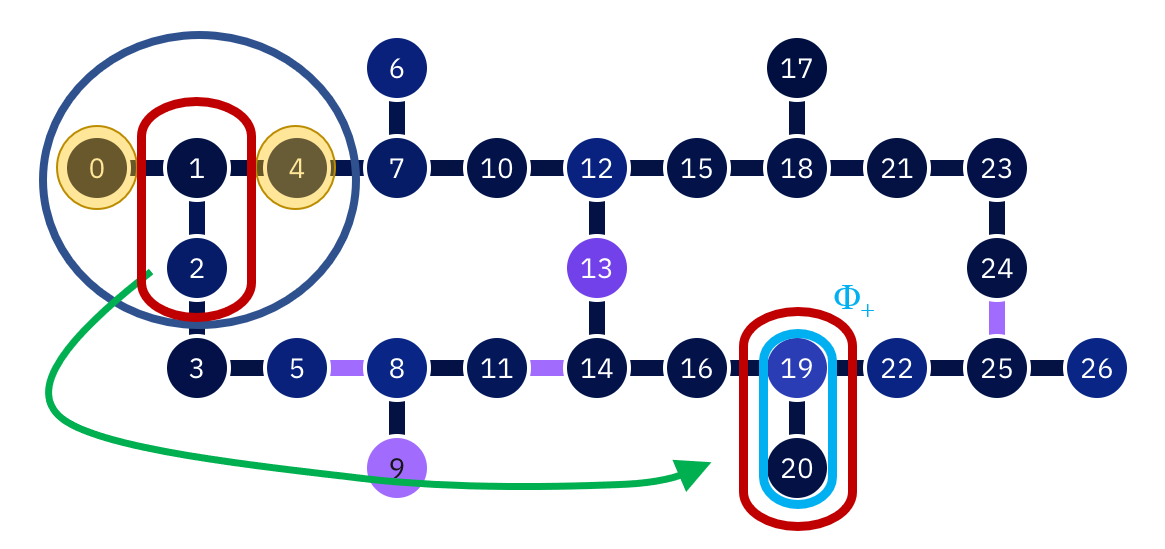}
    \caption{Borderline-quantumness of the cat-state protocol, for $M=4$ and
$J=2$.}
    \label{fig:success-cat-state-protocol}
\end{figure}

The cat state protocol 
$\{M=4;J\}$ (generalizing the Bell-state transfer, where Alice transmits $J$ qubits out of her substructure of M qubits)  
shows both quantumness and classicality depending on 
the parameter $J$ already for a non-maximal distance:
It failed to show quantumness for $\{M=4;J=3\}$ to distance $L=6$,
yielding a fidelity 0.47, see 
Figure~\ref{fig:failed-cat-state-protocol}.  
However, it shows borderline-successful quantumness, 
for $\{M=4;J=2\}$ at the same distance
$L=6$, see Figure~\ref{fig:success-cat-state-protocol}, 
yielding a fidelity
of 0.53, slightly above the quantumness threshold. 

\subsection{Fidelity of ancilla qubits}~\label{ssec:checking-ancillas} 
\begin{figure}[H]
    \centering
    \includegraphics[width=0.75\linewidth]{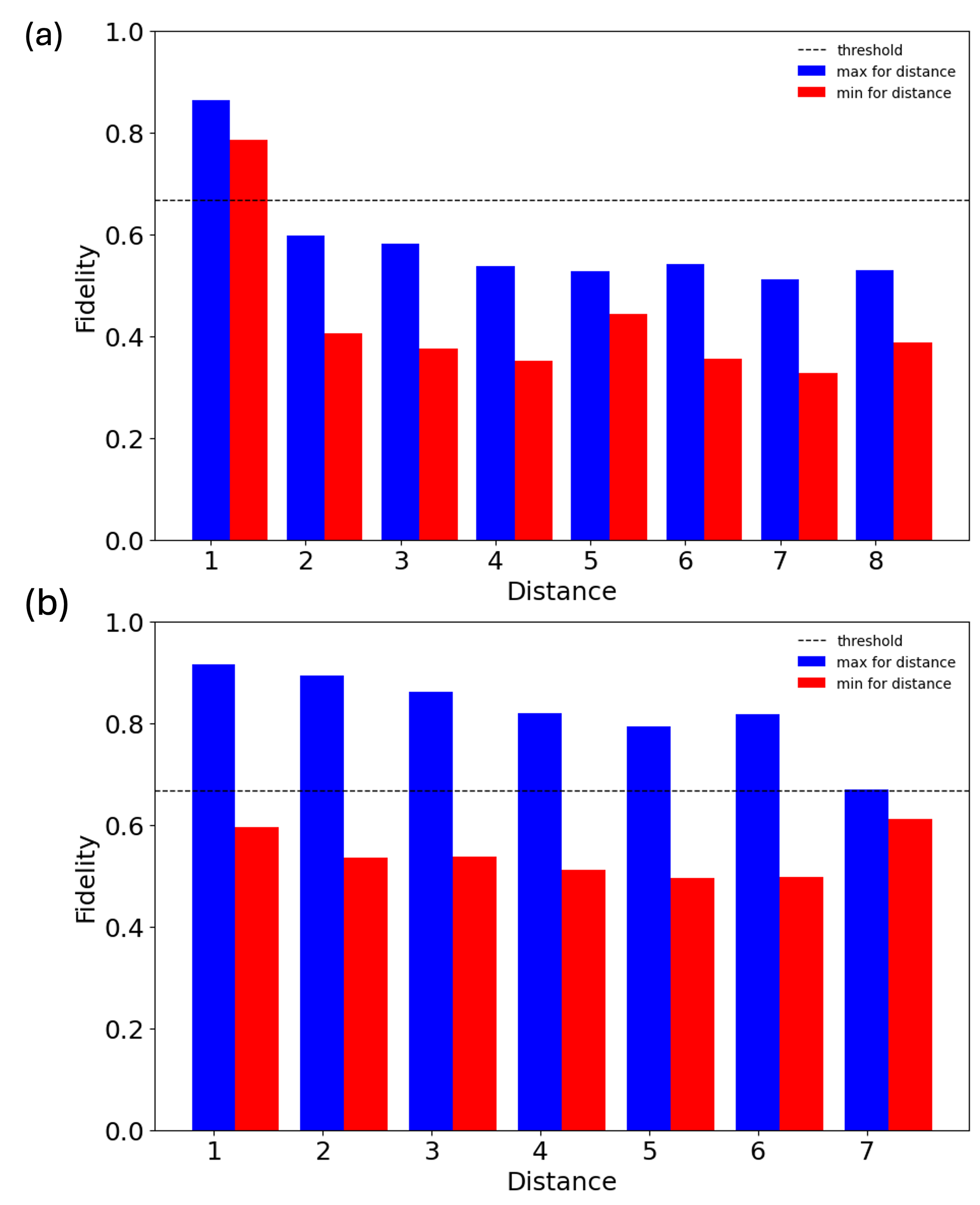}

    \caption{\textit{ibm\_melbourne} device results on do-nothing protocol with ancillas initialized to a random state.
(a) results from real device 
(b) results from a simulation using the device's noise model.}
    \label{fig:Melbourne_do_nothing_ancilla}
\end{figure}

The presented protocols measured only the fidelity of the work qubits. Here we present a variation to the protocols, taking into account also the protocols' effect on the ancilla qubits. As presented in Section~\ref{sec:protocols}, in these variants the ancilla qubits are also initialized to a random state and measured at the end: after applying the random initialization inverse. The fidelity of each qubit is acquired separately after marginalizing the resulting counts.
We tested this variant of the do-nothing protocol on the Melbourne device - both in simulation and on the real device. We tested all of the possible 
shortest linear subsets of the device, while taking into account only shortest paths between qubits.
Both the work qubit and all of the ancillas should pass the quantumness fidelity, therefore we present the worst qubit fidelity as the fidelity used to determine the distance in which the quantumness of the protocol remains. The results are shown in Figure~\ref{fig:Melbourne_do_nothing_ancilla}. In this variant, the results are determined by the worst qubit in the path, therefore the fidelities tend to be lower in comparison to the original variant.

\subsection{IBM-Q's Eagle}~\label{ssec:res-Eagle} 
\begin{figure}[H]
    \centering
    \includegraphics[width=0.5\linewidth]{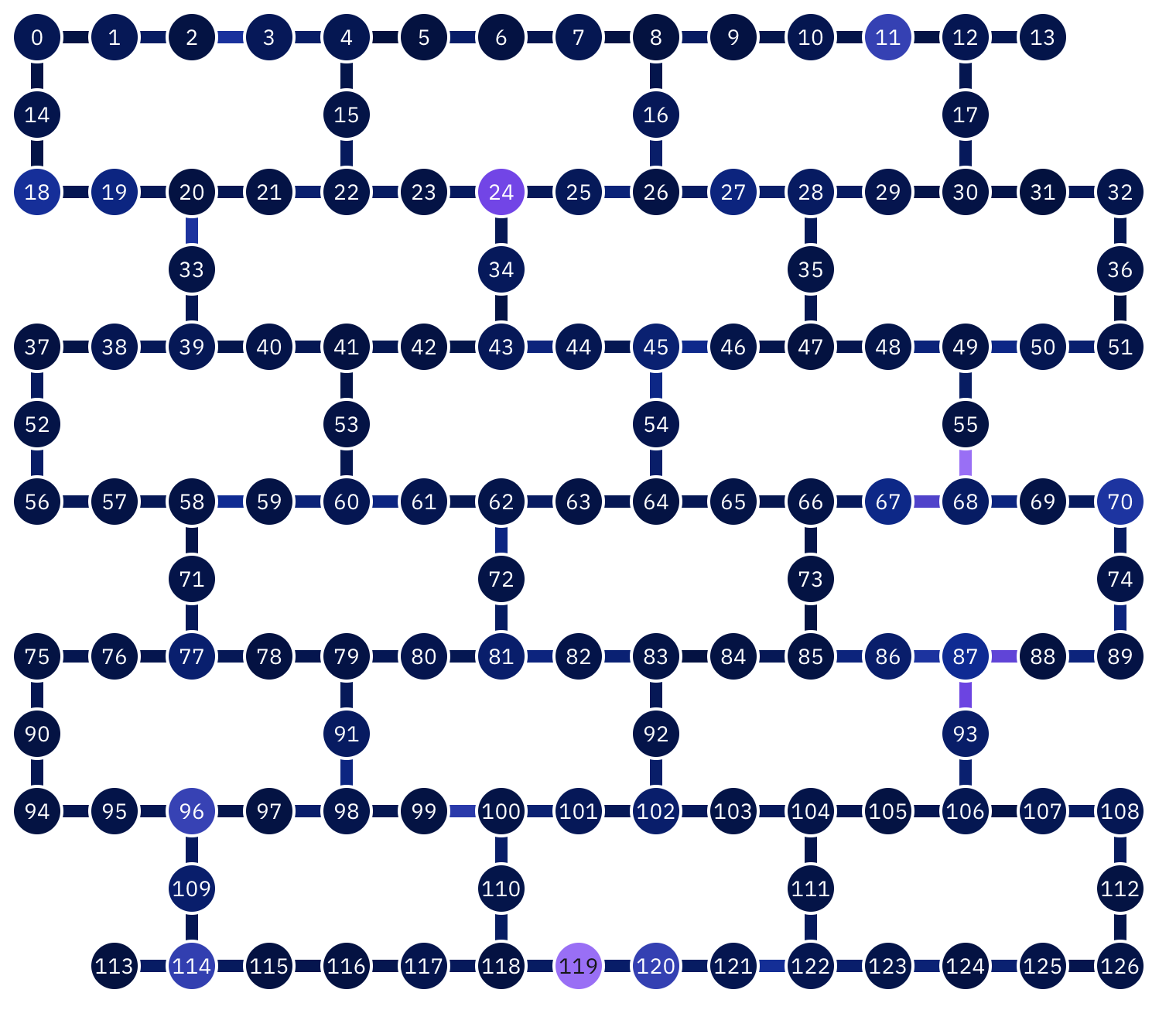}
    \caption{IBM-Q's Eagle connectivity map with 127 qubits.}
    \label{fig:Eagle}
\end{figure}

In this section we present our results on \textit{ibm\_brisabne}, 
one of IBM's Eagle devices, facilitating 127 qubits, 
both from simulation and real device. 
The device connectivity can be seen in Figure~\ref{fig:Eagle}.  
On this device we checked the do-nothing protocol for various distances. 
Because of the large amount of possible paths in the device, 
we limited our check only to circuits starting at $q_0$.
For each path we used $10000$ shots to acquire the fidelity 
of the protocol. Using these results, 
we obtained, for each distance from $q_0$, 
the maximal and minimal fidelities achieved in the various paths. 

\begin{figure}[H]
    \centering
    \includegraphics[width=0.75\linewidth]{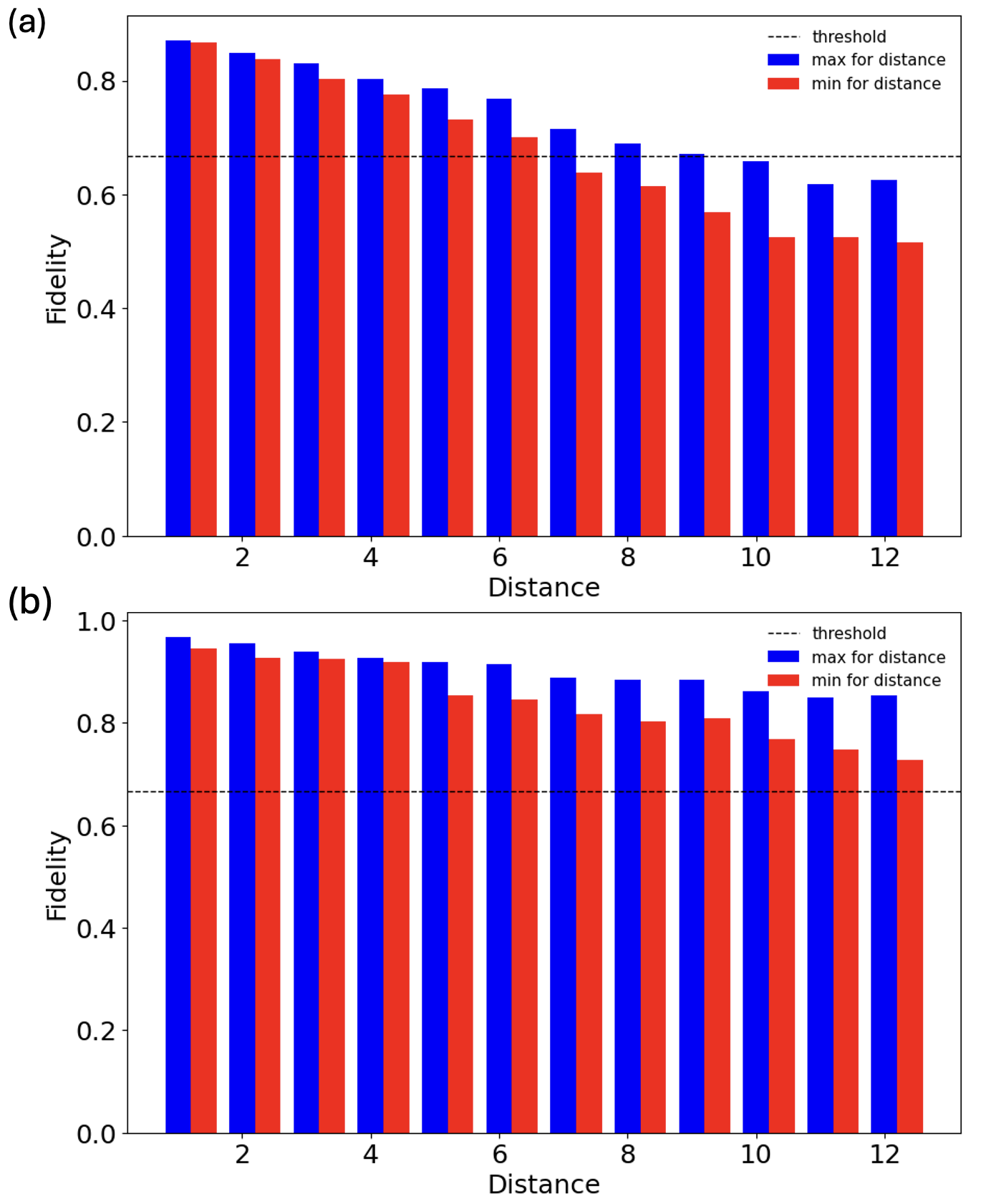}

    \caption{\textit{ibm\_brisabne} device results on do-nothing protocol up to
distance 12, on paths starting at $q_0$. 
(a) results from real device 
(b) results from a simulation using the device's noise model.}
    \label{fig:Brisbane_do_nothing_compare}
\end{figure}

\begin{figure}[H]
    \centering
   \includegraphics[width=0.6\linewidth]{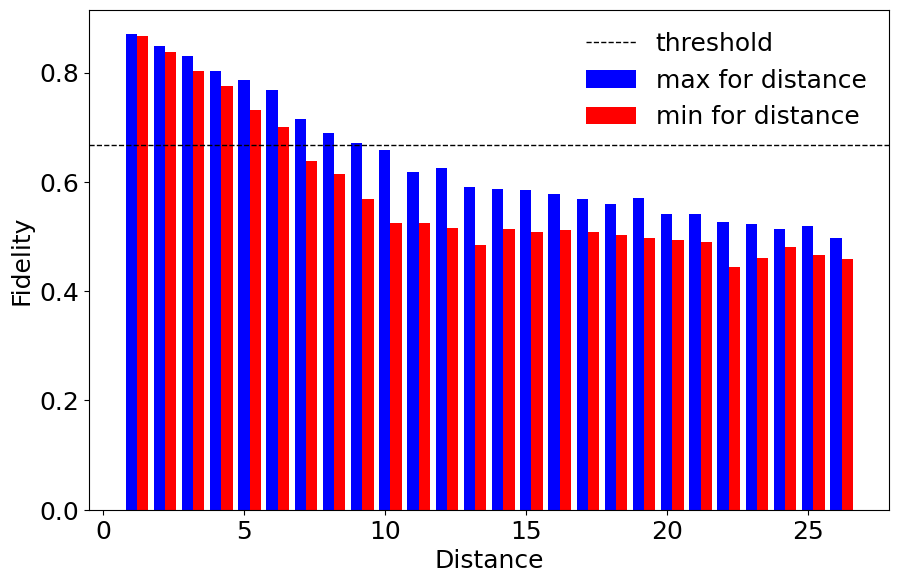}
    
\caption{\textit{ibm\_brisabne} real device results on do-nothing protocol on shortest paths starting at qubit $0$, up to any distance.} 
    \label{fig:Brisbane_real_do_nothing}
\end{figure}

In Figure~\ref{fig:Brisbane_do_nothing_compare} we compare the real device and the simulation for distances up to 12. Limiting the length to 12 allows us to also compare Brisbane to Kolkata. In Figure~\ref{fig:Brisbane_real_do_nothing} we show the results on the real device for any distance of shortest paths. 
The real device's maximal fidelity crosses the quantumness threshold 
at distance 10, meaning that for each distance up until 
distance 9 there exists a path which remains quantum on 
the do-nothing protocol. 
The real device's minimal fidelity crosses the quantumness threshold 
at distance 7, meaning that for each distance up until distance 6 
every path remains quantum on the do-nothing protocol.

In order to simulate the device, we used the noise model of the device 
supplied by IBM, retrieved from the function \textit{from\_backend} 
of the device's backend class in Qiskit package. 
It can be seen that the simulation shows better results, 
as the noise model takes into account some noise sources, given by Qiskit~\cite{qiskit_aer_noise_tutorial}, but not all of the possible noise sources 
which exists in the real device, such as crosstalk~\cite{crosstalk} and leakage~\cite{leakage} noises.

The results show that for distances up to $L = 10$ the decrease in fidelity as a function of path length is noticeably steeper in the real device compared to our simulation. Also, the real Brisbane gives worse fidelity results compared to real Kolkata, which is a much older chip model, see Figure~\ref{fig:Kolkata_do_nothing_real_device}.

\subsection{Simulated ion trap devices}~\label{ssec:res-fake-IonQ} 

An important feature of our protocols is their relevance to 
any circuit-based quantum computer no matter what the 
underlying hardware implementing the qubits is.
However, the connectivity map certainly influences the search
for optimal paths within a chip or a subschip.

\begin{figure}[H]
    \centering
    \includegraphics[width=0.5\linewidth]{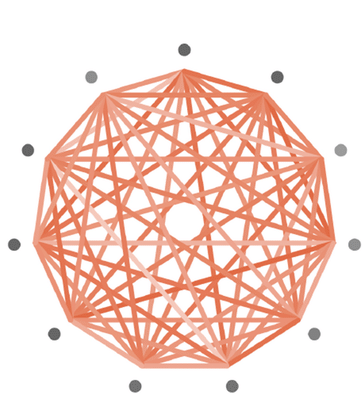}
    \caption{Ion-trap-11-qubits}
    \label{fig:Ion-trap-11}
\end{figure}

We present the results of some of the protocols on a simulation 
of two IonQ's trapped ion devices, Harmony of 11 qubits and Aria 
of 25 qubits. If one ignores the phonon needed for two-qubit gates, 
the connectivity map of the IonQ trapped ion devices becomes 
a complete graph (all-to-all connectivity), see Figure \ref{fig:Ion-trap-11} for the 
connectivity map of the Harmony device (11 qubits). 
With such connectivity, the maximum distance between any two qubits is 1, 
when considering only shortest paths. Therefore in order to find 
the amount of effective qubits using the protocols vector, 
one should choose protocols that use several qubits and paths,
and preferably also the generalized protocols, in order 
to increase the number of used qubits and paths.

Here we only ran the Bell-state transfer and entanglement swapping protocols. 
We ran them on the online simulators supplied by IonQ, 
featuring rather naive noise models of the Harmony and Aria devices, with identical noise parameters for all qubits.

The fidelities of the Bell-state transfer protocol (using 4 qubits 
and two paths) were 0.4441 for the Harmony device simulation 
and 0.9075 for the Aria device simulation.
The fidelities of the entanglement swapping protocol (using 6 qubits and 
two paths) were 0.3688 for the Harmony device simulation and 0.8854 
for the Aria device simulation.

Because of the all-to-all connectivity, a single swap gate was used 
for each sent qubit in order to move the qubit between Alice and Bob. 
Aria's hardware is a newer generation of IonQ's hardware, 
showing significant improvement in noise reduction. We leave checking 
the real devices for future research.

\section{Discussion and Acknowledgements}~\label{sec:Conclusion} 

We designed protocols for benchmarking quantum computers, and we presented
various variants and many examples, mainly on simulated chips. 

We focused on the case of a single path connecting Alice and Bob, on IBM
real and simulated (fake) devices, 
with only very few examples of more than a single path on Ion-Q
simulated devices. We provided cases for
comparing real and simulated devices. We usually ignored the less-involved
ancilla qubits, but we also measured them in one of our experiments.

We did not employ intermediate measurements, but we plan to do so in future
work, as this option presently exists.

A historical note: 
Already around 2018 we (T.M.~along with two of his students, 
Chen Mechel and Rotem Liss) 
designed and ran a few initial teleportation and entanglement-swapping 
experiments on a chip that resembles Melbourne 
but did not exist anymore when we re-initiated the project.
T.M.~presented this research in several scientific meetings, wherein discussions with colleagues and peers contributed research directions explored in this work. 
A document and a presentation containing the data and 
results from 2018 can be provided upon request.

The project was then re-initiated and much improved and extended --- 
to reach the current stage as presented in this paper.

Acknowledgements: D.M. and T.M. thank the Helen Diller Quantum
Center at the Technion for their generous Support. T.M. and Y.W. thank the
Quantum Computing Consortium of Israel Innovation Authority for financial
support.
T.M. and Y.W. thank the Israeli MOD for their support at the early stages 
of this work. 
We all thank Chen Mechel and Rotem Liss for their contribution to the very 
early stage of experimenting teleportation and entanglement swapping. 

\bibliographystyle{plain}
\bibliography{Bench1-18May25}

\end{document}